\documentclass[final,5p,times,twocolumn]{elsarticle}

\usepackage{amssymb}
\usepackage{amsmath}
\usepackage[utf8]{inputenc}
\usepackage[T1]{fontenc}
\usepackage{lineno}

\usepackage{hyperref}
\hypersetup{
    colorlinks=true,
    linkcolor=blue}

\usepackage{array,tabularx,caption}    
\journal{NIM A}

\begin{document}
\begin{frontmatter}



\title{Advances in photocathode development for PICOSEC Micromegas precise-timing detectors}

\author[inst1]{M. Lisowska\corref{cor1}}
\cortext[cor1]{Corresponding author.}
\ead{marta.lisowska@cern.ch}
\author[inst1,inst2]{F. Guerra}
\author[inst1,inst29,inst31]{A. Gurpinar}
\author[inst17]{D. Zavazieva}
\author[inst3]{R. Aleksan}
%\author[inst4,inst5]{Y. Angelis}
\author[inst3]{S. Aune}
\author[inst6]{J. Bortfeldt}
\author[inst30]{A. Breskin}
\author[inst1]{F.M. Brunbauer}
\author[inst7,inst8]{M. Brunoldi}
%\author[inst4,inst5]{E. Chatzianagnostou}
\author[inst9]{J. Datta}
%\author[inst10]{K. Dehmelt}
\author[inst11]{G. Fanourakis}
\author[inst1]{S. Ferry}
%\author[inst7,inst8]{D. Fiorina\fnref{fn1}}
\author[inst1,inst12]{K. J. Floethner}
\author[inst13]{M. Gallinaro}
\author[inst14]{F. Garcia}
\author[inst3]{I. Giomataris}
%\author[inst10]{K. Gnanvo}
%\author[inst3]{F.J. Iguaz\fnref{fn2}}
\author[inst1]{D. Janssens}
\author[inst1,inst27]{E.~Jelinkova}
\author[inst3]{A. Kallitsopoulou}
\author[inst4]{I. Karakoulias}
\author[inst16]{M. Kovacic}
%\author[inst10]{B. Kross}
%\author[inst17]{C.C. Lai}
\author[inst3]{P. Legou}
\author[inst18]{J. Liu}
\author[inst12,inst19]{M. Lupberger}
%\author[inst1,inst4]{I. Maniatis\fnref{fn3}}
\author[inst1]{D.J.G. Marques}
%\author[inst10]{J. McKisson}
\author[inst18]{Y. Meng}
\author[inst1,inst19]{H. Muller}
\author[inst1]{R. De Oliveira}
\author[inst1]{E. Oliveri}
%\author[inst1,inst20]{G. Orlandini}
%\author[inst10]{A. Pandey}
\author[inst3]{T. Papaevangelou}
\author[inst21]{M. Pomorski}
%\author[inst1,inst22]{M. Robert}
\author[inst1]{L. Ropelewski}
\author[inst4,inst5]{D. Sampsonidis} 
%\author[inst1]{L. Scharenberg}
\author[inst1]{T. Schneider}
\author[inst1,inst28]{B.~Schoenfelder}
\author[inst21]{E. Scorsone}
%\author[inst9]{N. Shankman}
%\author[inst1,inst3]{L. Sohl\fnref{fn4}}
\author[inst1]{M. van Stenis}
\author[inst23]{Y. Tsipolitis}
\author[inst4,inst5]{S. Tzamarias}
\author[inst24]{A. Utrobicic}
\author[inst7,inst8]{I. Vai}
\author[inst1]{R. Veenhof}
\author[inst1]{L. Viezzi}
\author[inst7,inst8]{P.~Vitulo}
%\author[inst1,inst25]{C. Volpato}
\author[inst18]{X. Wang}
\author[inst1,inst26]{S. White}
%\author[inst10]{W. Xi}
\author[inst18]{Z. Zhang}
\author[inst18]{Y. Zhou}

\affiliation[inst1]{organization={European Organization for Nuclear Research (CERN), 1211 Geneve 23, Switzerland}}

\affiliation[inst2]{organization={Department of Physics, University of Turin, 10125 Torino, Italy}}

\affiliation[inst29]{organization={Department of Chemistry, Middle East Technical University, 06800 Ankara, Turkey}}

\affiliation[inst31]{organization={METU MEMS Center, Middle East Technical University, 06800 Ankara, Turkey}}

\affiliation[inst17]{organization={Technion - Israel Institute of  Technology, Haifa, 3200003, Israel}}

\affiliation[inst3]{organization={IRFU, CEA, Université Paris-Saclay, F-91191 Gif-sur-Yvette, France}}

\affiliation[inst6]{organization={Department for Medical Physics, Ludwig Maximilian University of Munich, Am Coulombwall 1, 85748 Garching, Germany}}

\affiliation[inst30]{organization={Weizmann Institute of Science, Herzl St 234, Rehovot, Israel}}
 
\affiliation[inst7]{organization={Dipartimento di Fisica, Università di Pavia, Via Bassi 6, 27100 Pavia, Italy}}

\affiliation[inst8]{organization={INFN Sezione di Pavia, Via Bassi 6, 27100 Pavia, Italy}}

\affiliation[inst9]{organization={Department of Physics and Astronomy, Stony Brook University, Stony Brook, NY 11794-3800, USA}}

\affiliation[inst11]{organization={Institute of Nuclear and Particle Physics, NCSR Demokritos, GR-15341 Agia Paraskevi, Attiki, Greece}}

\affiliation[inst12]{organization={Helmholtz-Institut für Strahlen- und Kernphysik, University of Bonn, Nußallee 14–16, 53115 Bonn, Germany}}

\affiliation[inst13]{organization={Laboratório de Instrumentacão e Física Experimental de Partículas, Lisbon, Portugal}}

\affiliation[inst14]{organization={Helsinki Institute of Physics, University of Helsinki, FI-00014 Helsinki, Finland}}

%\affiliation[inst10]{organization={Jefferson Lab, 12000 Jefferson Avenue, Newport News, VA 23606, USA}}

\affiliation[inst27]{organization={Czech Technical University in Prague, Jugoslávských partyzánů 1580/3, 160 00 Praha 6-Dejvice, Czechia}}

\affiliation[inst4]{organization={Department of Physics, Aristotle University of Thessaloniki, University Campus, GR-54124, Thessaloniki, Greece}}

\affiliation[inst16]{organization={University of Zagreb, Faculty of Electrical Engineering and Computing, 10000 Zagreb, Croatia}}

%\affiliation[inst17]{organization={European Spallation Source (ESS), Partikelgatan 2, 224 84 Lund, Sweden}}

\affiliation[inst18]{organization={State Key Laboratory of Particle Detection and Electronics, University of Science and Technology of China, Hefei 230026, China}}

\affiliation[inst19]{organization={Physikalisches Institut, University of Bonn, Nußallee 12, 53115 Bonn, Germany}}

%\affiliation[inst20]{organization={Friedrich-Alexander-Universität Erlangen-Nürnberg, Schloßplatz 4, 91054 Erlangen, Germany}}

\affiliation[inst21]{organization={CEA-LIST, Diamond Sensors Laboratory, CEA Saclay, F-91191 Gif-sur-Yvette, France}}

\affiliation[inst5]{organization={Center for Interdisciplinary Research and Innovation (CIRI-AUTH), Thessaloniki 57001, Greece}}

%\affiliation[inst22]{organization={Queen’s University, Kingston, Ontario, Canada}}

\affiliation[inst28]{organization={Fakultät für Physik, Technische Universitaet Wien, Karlsplatz 13, 1040 Vienna, Austria}}

\affiliation[inst23]{organization={National Technical University of Athens, Athens, Greece}}

\affiliation[inst24]{organization={Ruđer Bošković Institute, Bijenička cesta 54., 10 000 Zagreb, Croatia}}

%\affiliation[inst25]{organization={Department of Physics and Astronomy, University of Florence, Via Giovanni Sansone 1, 50019 Sesto Fiorentino, Italy}}

\affiliation[inst26]{organization={University of Virginia, USA}}

%\fntext[fn1]{Now at Gran Sasso Science Institute, Viale F. Crispi, 7 67100 L'Aquila, Italy.}
%\fntext[fn2]{Now at SOLEIL Synchrotron, L’Orme des Merisiers, Départementale 128, 91190 Saint-Aubin, France.}
%\fntext[fn3]{Now at Department of Particle Physics and Astronomy, Weizmann Institute of Science, Hrzl st. 234, Rehovot, 7610001, Israel.}
%\fntext[fn4]{Now at TÜV NORD EnSys GmbH \& Co. KG.}

\begin{abstract}

The PICOSEC Micromegas detector is a~precise-timing gaseous detector that combines a~Cherenkov radiator, a~semi-transparent photocathode and a~Micromegas amplification stage, targeting time resolutions of tens of picoseconds for minimum ionising particles (MIPs).
Initial single-pad prototypes achieved time resolutions of $\sigma<25$\,ps, demonstrating strong potential for High Energy Physics (HEP) applications and beyond.
The objective of this paper is a~comprehensive characterisation of photocathodes, with a~strong focus on robust materials while preserving excellent timing performance.
The study includes laboratory measurements of optical and resistive properties, along with beam tests using 150\,GeV/$c$ muons to evaluate the time resolution and photoelectron yield for various photocathodes.
The best performance was obtained by a~5\,nm Cesium Iodide (CsI) photocathode, reaching $\sigma = 10.9 \pm 0.3$\,ps with more than 30~extracted photoelectrons, representing the most precise time resolution achieved by PICOSEC Micromegas to date.
Metallic and carbon-based photocathodes, including Titanium (Ti), Boron Carbide (B$_4$C) and Diamond-Like Carbon (DLC), were also tested, with Ti and B$_4$C emerging as the most promising alternatives, achieving $\sigma \approx 30$\,ps with about 5 extracted photoelectrons.
These results demonstrate that improved robustness can be achieved while maintaining excellent time resolution, supporting the feasibility of using the PICOSEC Micromegas concept in future experiments.

\end{abstract}

%Graphical abstract

%Research highlights
%\item The excellent timing performance of the single-channel proof of concept was not only transferred to the 100-channel prototype, but even improved to $\sigma$~=~18 ps, making the PICOSEC Micromegas detector more suitable for large-area experiments in need of detectors with high time resolutions.

\begin{keyword}
Gaseous detectors \sep Micromegas \sep Photocathodes \sep Timing resolution
\end{keyword}

\end{frontmatter}

%\begin{linenumbers}

%\linenumbers

%\section{Sample Section Title}
%\label{sec:sample1}

%%       \citet{<label>} ==> Jones et al. [21]
%%       \citep{<label>} ==> [21]
%%

\begin{figure*}[!t]
\begin{center}
\includegraphics[width=17cm]{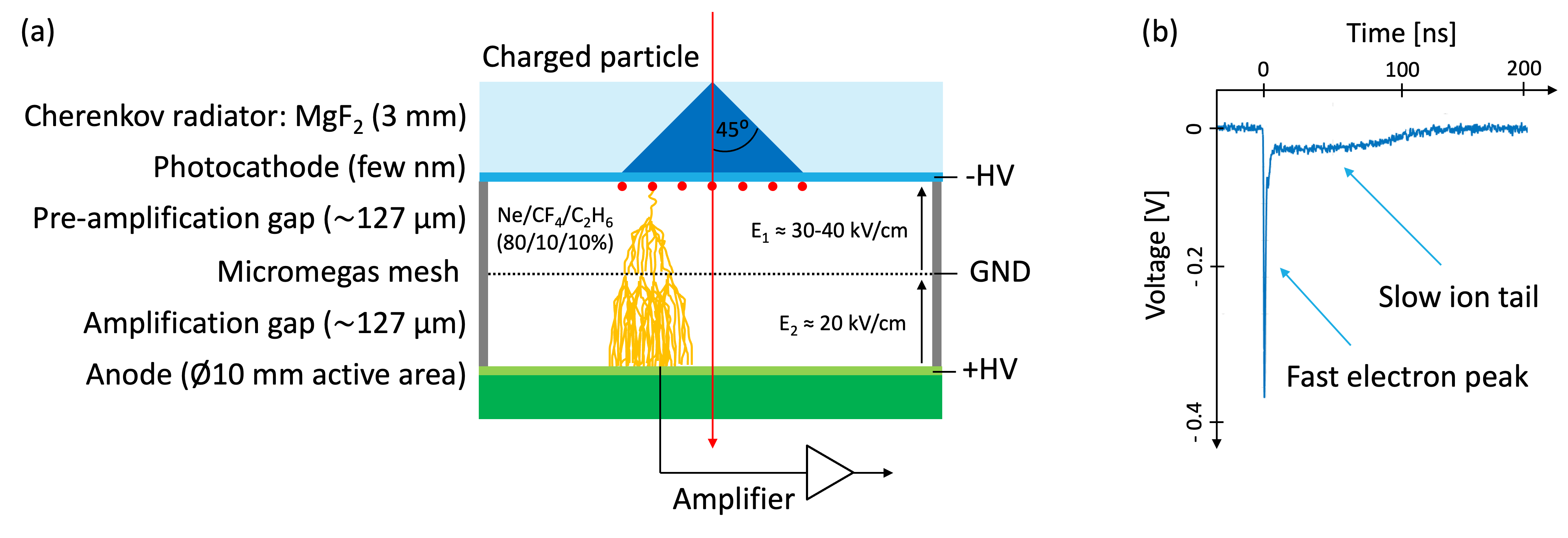}
\end{center}
\caption{(a) PICOSEC Micromegas detection concept: a~charged particle traversing a~Cherenkov radiator generates UV photons, a~fraction of which is converted into electrons at a~photocathode.
The electrons undergo two-stage gas multiplication and the resulting charge motion in the amplification gap induces a~signal on the anode.
%The measurements presented in this paper were performed using a single-pad metallic detector with a 10 mm diameter active area.
A custom current-sensitive amplifier (5--650~MHz bandwidth, 38.5~dB gain, 44~$\Omega$, ENC $\sim$1100-1700~$\mathrm{e^-}$) was used to amplify the signal.
Note that the figure is not drawn to scale.  
(b)~Typical PICOSEC Micromegas waveform: The signal features a~fast electron peak followed by a~slow ion tail~\cite{LisowskaPhDThesis}.}
\label{PicosecDetectionConcept}
\end{figure*}

~
\newline
~
\newline

\section{Introduction}
\label{sec:sample1}

%\begin{linenumbers}

The strong interest in precise-timing detectors is driven by the demanding conditions of future HEP experiments and beyond~\cite{ECFA}. Sub-nanosecond resolution is essential for separating closely spaced events, improving track reconstruction and enabling particle identification via time-of-flight (ToF) measurements. To address the high pileup expected at the High Luminosity Large Hadron Collider (HL-LHC), experiments are upgrading to precise-timing detectors such as the CMS Minimum Ionising Particles Timing Detector (MTD)~\cite{CMS_MTD}. The MTD is expected to provide time resolutions of 30-40 ps initially, degrading to 50–60 ps over HL-LHC operation due to radiation damage.

The need for precise timing extends beyond HEP. In positron emission tomography (PET), ToF information from photon pairs produced in electron–positron annihilation is used to localise tumours more precisely. Current scintillator-based systems achieve time resolutions of $\mathcal{O}(100)$\,ps, corresponding to spatial precision of about 1 cm~\cite{PETToF}. Achieving millimetre-scale resolution instead requires time resolutions of $\mathcal{O}(10)$\,ps for photons. Continuous advances in detector technology therefore aim to further improve timing performance.

The PICOSEC Micromegas project~\cite{firstPicosecPaper} aims to develop a~robust multi-channel gaseous detector with a~target time resolution of $\mathcal{O}(10)$\,ps for MIPs. Early single-pad prototypes with CsI photocathodes achieved time resolutions $\sigma~<~25$\,ps~\cite{firstPicosecPaper}, motivating further improvements in detector design \cite{UtrobicicSinglePad,UtrobicicResistiveSinglePad}, stability and robustness \cite{LukasPhDThesis,myMScThesis,XuPhotocathodes,LisowskaPhDThesis,myPhotocathodes,GuerraSummer}, as well as scalability to larger areas~\cite{multipadPicosecPaper,NDIP,AntonijaMPGD,MengMPGD,KallitsopoulouPhDThesis,Kallitsopoulou7pad,Kallitsopoulou96pad}.

While devices such as micro-channel plate photomultiplier tubes (MCP-PMTs) and other detectors with Cherenkov light-producing windows can achieve excellent timing performance, down to the few-picosecond level~\cite{LukasPhDThesis}, their capabilities are typically limited to small active areas and they are associated with high costs. In contrast, PICOSEC Micromegas detectors provide competitive timing performance over significantly larger areas, with current prototypes covering 10×10 cm$^2$ and 20×20 cm$^2$ while maintaining time resolutions at the level of $\sigma~\approx~20-25$\,ps \cite{NDIP,AntonijaMPGD,MengMPGD}. This scalability makes PICOSEC a~promising solution for large-area precise-timing applications.

Traditional photocathode materials such as CsI offer high quantum efficiency (QE) and strong ultraviolet (UV) sensitivity, but suffer from vulnerability to ion backflow and discharges, as well as humidity. These limitations motivate the development of more robust alternatives. Although preliminary studies of other materials have been performed~\cite{LukasPhDThesis,myMScThesis,XuPhotocathodes,LisowskaPhDThesis,myPhotocathodes,GuerraSummer},
a~comprehensive characterisation of metallic and carbon-based photocathodes - covering both time resolution and photoelectron yield - remains lacking. This work addresses this gap by developing more robust photocathodes while maintaining excellent timing performance.

Within the scope of this paper, the characterisation of four photocathode materials - CsI, Ti, B$_4$C and DLC - and their performance as candidates for PICOSEC Micromegas detectors are reported.
The samples were manufactured at the Thin Film and Glass (TFG) and Micro-Pattern Technologies (MPT) workshops at the European Organization for Nuclear Research (CERN). The measurements were carried out during two test beam campaigns in July and November~2025. During the November campaign, some previously studied samples were remeasured to assess the reproducibility of earlier results and to determine the photoelectron yield, which had not been evaluated in earlier test beams.

\section{PICOSEC Micromegas detection concept}
\label{sec:2}

The PICOSEC Micromegas detection concept~\cite{firstPicosecPaper} is illustrated in Fig.~\ref{PicosecDetectionConcept}a. 
The detector reduces time jitter from primary gas ionisation by using a~Cherenkov radiator. 
A charged particle traversing the Cherenkov radiator produces UV photons.
For 150 GeV/c muons crossing a~3 mm MgF$_2$ crystal, Cherenkov radiation corresponding to approximately 230 UV photons per track is emitted at an angle of about 45$^\circ$.
A fraction of the photons is converted into electrons at a~semi-transparent photocathode, determined by its QE.
Since all electrons are created at the same surface, the uncertainty in ionisation location is minimised, improving the timing precision.

The detector operates in a~Ne/C$_2$H$_6$/CF$_4$ (80/10/10\%, volumetric) gas mixture ~\cite{COMPASS}. 
A calendered (i.e. mechanically compressed) stainless-steel woven mesh (18~$\mu$m wires, 45~$\mu$m opening, 30 $\mu$m thickness, 50\% optical transparency \cite{firstPicosecPaper}) protects the photocathode from ion backflow and suppresses discharges.
A copper-clad polyimide spacer defines the pre-amplification gap. 
Electric fields of $\sim$30-40\,kV/cm in the pre-amplification gap and $\sim$20\,kV/cm in the amplification gap yield total gains of $10^{5}$-$10^{6}$. 

The motion of electrons toward the anode and ions toward the mesh in the amplification gap induces a~signal, which is amplified and digitised.
A typical waveform, showing a~fast electron peak followed by a~slow ion tail, is shown in Fig.~\ref{PicosecDetectionConcept}b.
The leading edge of the electron peak defines the Signal Arrival Time (SAT).
A~detailed description of the detection concept is provided in~\cite{LisowskaPhDThesis}.

\begin{figure}[!t]
\begin{center}
\includegraphics[width=\columnwidth]{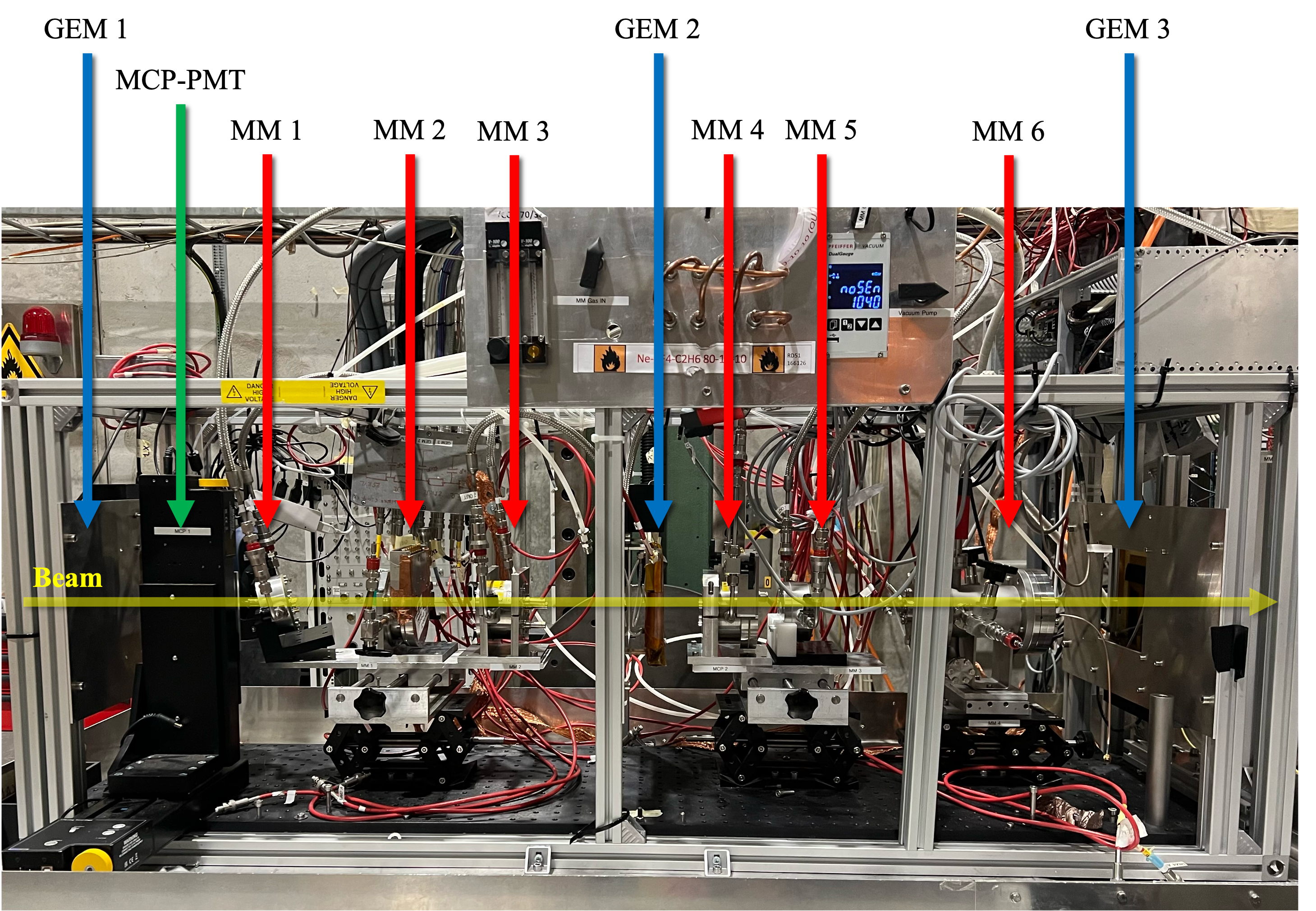}
\end{center}
\caption{Beam telescope setup including three triple-GEM detectors for particle tracking, an MCP-PMT as timing reference and DAQ trigger, together with PICOSEC Micromegas prototypes for testing~\cite{LisowskaPhDThesis}.}
\label{Telescope}
\end{figure}

\begin{figure*}[!t]
\begin{center}
\includegraphics[width=\textwidth]{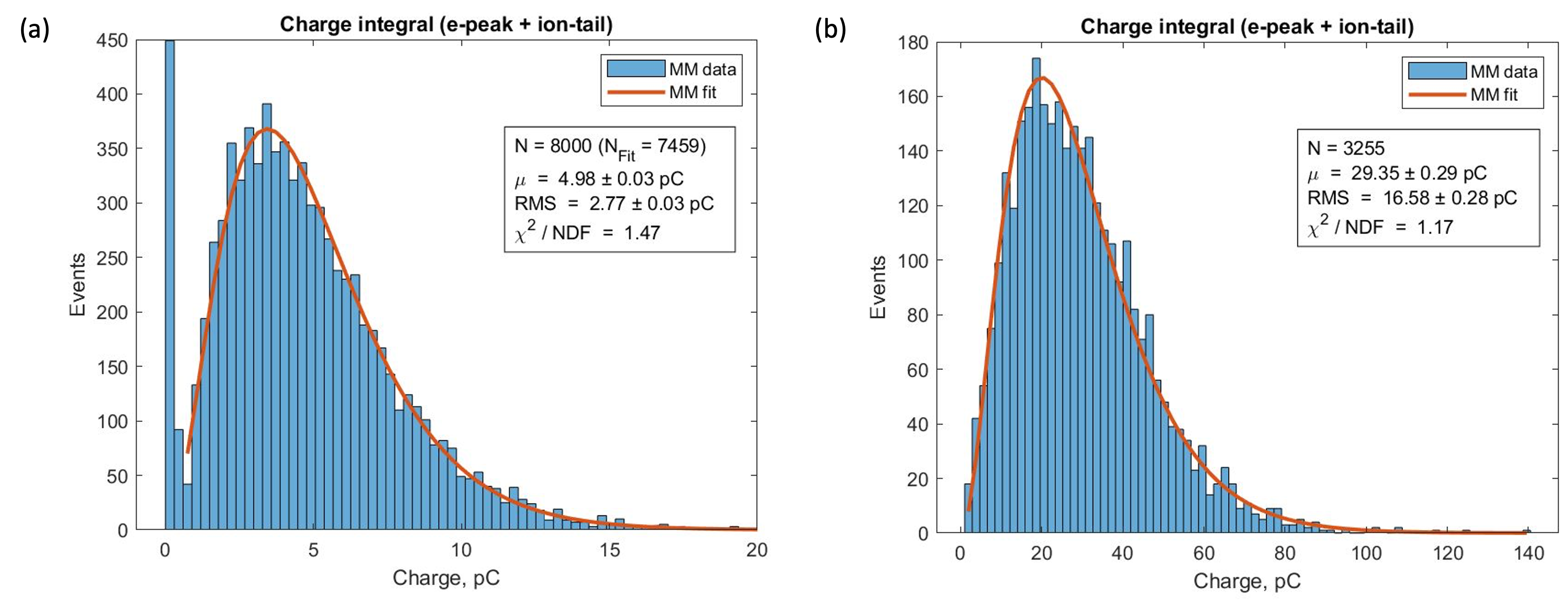}
\end{center}
\caption{Examples of the charge spectra for (a) SPE and (b) MIP signals, each fitted with a~Pólya distribution, for a~5 nm B$_4$C photocathode. The $N_{\text{PE}}$ is obtained as the ratio of the mean MIP charge to the mean SPE charge. The low-charge peak in the SPE histogram corresponds to noise and is excluded from the calculation of the mean signal charge.}
\label{PolyaNPE}
\end{figure*}

\section{Experimental methodology}
\label{sec:3}

%I think this section is ready

Photocathode optical measurements were performed using the ASSET (A Small Sample Evaporation Test) setup \cite{myMScThesis,ALICE_RICH,ALICE_CsI}, developed for quantitative studies of QE, transparency and ageing. ASSET provides tunable Vacuum-UltraViolet (VUV) illumination in the 120–200 nm range via a~deuterium lamp \cite{deuteriumLamp} and monochromator system \cite{Monochromator} and allowes operation in reflective and transmission modes under a~vacuum of 10$^{-6}$~mbar.
It uses calibrated CsI photomultiplier tubes (PMTs) to monitor light intensity for precise current-based QE and transparency extraction.
Transparency in the 200–800 nm range is measured with a~PerkinElmer Lambda 650 UV/VIS spectrophotometer \cite{SpectrometerTransparency}.
ASSET also enables controlled degradation measurements by exposing samples to ion bombardment generated in an X-ray–irradiated gas chamber. A~complete technical description of the ASSET apparatus and measurement procedures is provided in previous publications~\cite{myMScThesis,LisowskaPhDThesis,myPhotocathodes}.
Scanning electron microscopy (SEM) was further employed to characterise and compare the surface morphology of selected photocathodes with varying thicknesses.

The thicknesses of the photocathode layers produced by physical vapour deposition at the TFG workshop were estimated using an Inficon deposition controller~\cite{Inficon}. This device uses a~quartz crystal sensor, in which the oscillation frequency of the crystal changes as material is deposited on its surface. By inputting material-specific parameters, such as density, the frequency shift can be converted into an estimate of the deposited layer thickness. In contrast, for photocathodes fabricated by magnetron sputtering at the MPT workshop, direct measurements of the film thicknesses were not available. Instead, thin films (on the order of a~few nanometres) were obtained by scaling the deposition time based on calibration runs performed with thicker films (tens of nanometres), which thicknesses were measured using a profilometer at the TFG workshop.

While, for the resistive-anode Micromegas, the optimal surface resistivity must balance two aspects - being low enough to minimise voltage drop under high-rate beam conditions and high enough to ensure stable operation without degrading the signal leading edge - for the photocathodes, surface resistivity measurements were conducted to further validate the thickness estimates.
%To further validate these estimates, surface resistivity measurements were conducted.
These were performed using a~picoammeter ~\cite{Keithley}. A~low voltage was applied across two conductive strips deposited on the substrates and the resulting photocathode resistivity was recorded. The results followed the expected trend, with thinner films exhibiting higher surface resistivity. Additionally, for photocathodes deposited on two different substrates, higher resistivity values indicated thinner layers on one substrate, which is likely related to reduced adhesion during the deposition process.

Particle beam campaigns were performed to measure the time resolutions of prototypes assembled in various configurations. 
The measurements used 150~GeV/c muon beams at the CERN SPS H4 beamline. 
The test setup included a~beam telescope providing triggering, timing and tracking capabilities, with an example configuration presented in Fig.~\ref{Telescope}.
Precise particle tracking was achieved using three triple Gas Electron Multiplier (GEM) detectors with spatial resolution below 80~$\mu$m. 
Signals from all GEM channels were shaped by APV25~\cite{APV25} front-end ASICs and digitised with the Scalable Readout System (SRS)~\cite{SRS}. 
The GEMs were operated in an Ar/CO$_2$ (70/30~\%) gas mixture at ambient pressure.
An MCP-PMT (Hamamatsu R3809U-58~\cite{MCP-PMT}) provided the timing reference and data acquisition (DAQ) trigger. 
The telescope setup allowed testing of multiple PICOSEC Micromegas prototypes simultaneously.

The overall time resolution of a~detector system, $\sigma_{\text{tot}}$, arises from several contributions and can be expressed as:
\begin{equation}
\sigma_{\text{tot}}^2~=~\sigma_{\text{MIP}}^2 + \sigma_{t_0}^2 + \sigma_{\text{e}}^2 + \dots
\label{eq:timeResolution}
\end{equation}
The first term, $\sigma_{\text{MIP}}$, is the PICOSEC Micromegas prototype resolution measured with MIPs and depends on the number of photoelectrons:
\begin{equation}
\sigma_{\text{MIP}}~=~\frac{\sigma_{\text{SPE}}}{\sqrt{N_{\text{PE}}}},
\label{eq:timeResolution2}
\end{equation}
where $\sigma_{\text{SPE}}$ is the single-photoelectron (SPE) resolution and $N_{\text{PE}}$ is the number of extracted photoelectrons. A~higher $N_{\text{PE}}$ leads to improved $\sigma_{\text{MIP}}$. 
The second term, $\sigma_{t_0}$, comes from the reference device, namely the MCP-PMT, while the third term, $\sigma_{\text{e}}$, arises from the electronics. The time resolutions reported in this paper include all these contributions.

%\subsubsection*{Time resolution}

A reference device with superior timing precision is required to quantify the PICOSEC Micromegas detectors time resolution. 
The previously mentioned MCP-PMT, with sub-5~ps resolution in the central region, serves this purpose and is aligned to the PICOSEC Micromegas prototypes to ensure particle passage through both devices. 
Signals are amplified using custom current-sensitive amplifiers (PICOSEC amplifiers hereafter) with a~bandwidth from 5 to 650 MHz, gain of 38.5~dB, input impedance of 44 $\Omega$ and an equivalent noise charge of about 1100–1700 electrons~\cite{AntonijaMPGD, RFamp}.
The amplified signals are recorded with a~LeCroy WR8104 oscilloscope~\cite{LeCroy}. 

Leading-edge discrimination with a~fixed threshold introduces time walk, where the SAT depends on amplitude - larger pulses cross earlier and appear sooner.
For PICOSEC Micromegas, this precision is insufficient, therefore software constant fraction discrimination (CFD) is used in the offline waveform analysis, defining the timestamp at a~fixed fraction of the peak amplitude.
This method assumes a~consistent signal shape.
To reduce noise, the leading edge is fitted with a~sigmoid function.
The time is extracted at the 20\% constant-fraction level \cite{firstPicosecPaper}. 
The SAT is calculated as the difference between the PICOSEC Micromegas and MCP-PMT timestamps. 
SAT distributions are fitted with a~double Gaussian:
\begin{equation}
f(\Delta t)~=~N \left[a \exp \left(-\frac{(\Delta t - \mu)^2}{2 \sigma^{2}_{\text{core}}}\right) + (1-a) \exp \left(-\frac{(\Delta t - \mu)^2}{2 \sigma^{2}_{\text{tail}}}\right) \right],
\label{eq:gauss}
\end{equation}
where \( N \) is a~scaling factor,  \( a~\in [0, 1] \) is a~weighting parameter, \( \mu \) is the mean of the distribution, \( \sigma_{\text{core}} \) is the standard deviation of the core Gaussian and \( \sigma_{\text{tail}} \) is the standard deviation of the Gaussian describing the tail.
The combined standard deviation $\sigma_{\text{comb}}$ of the overall distribution:
\begin{equation}
\sigma^{2}_{\text{comb}}~=~a~\sigma^{2}_{\text{core}} + (1-a)~ \sigma^{2}_{\text{tail}},
\label{eq:gauss_comb}
\end{equation}
represents the time resolution of the detector system.
Detector efficiency is defined as the fraction of events inducing a~signal registered by the PICOSEC Micromegas prototype compared to the MCP-PMT.

\begin{figure*}[!t]
\begin{center}
\includegraphics[width=\textwidth]{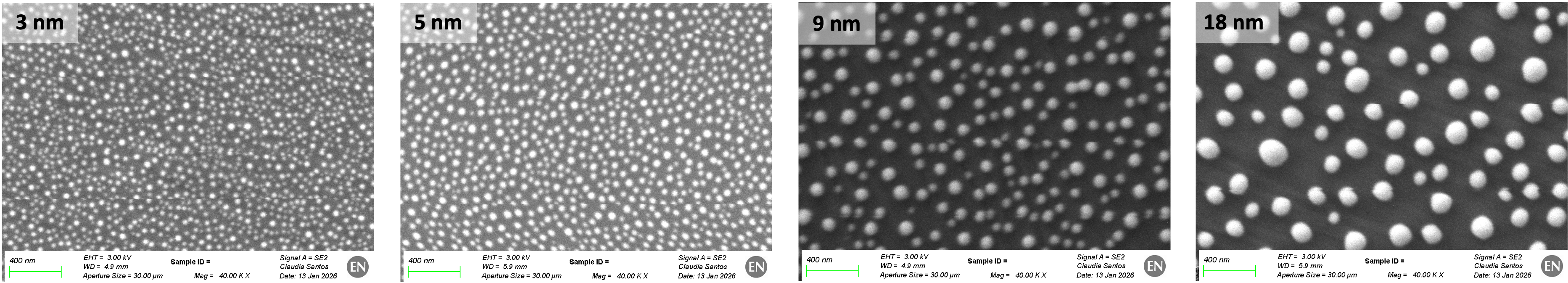}
\end{center}
\caption{SEM images of CsI photocathodes of different thicknesses deposited on the MgF$_2$ radiator with a 2.4~nm Ti interfacial layer in between, which serves as a~HV contact and mitigates charging-up effects. All samples were briefly exposed to air for a~total of no more than 2~minutes. The images show a~grain-like structure, with grain size increasing with layer thickness. The thinnest layer exhibits the most uniform morphology. The mismatch between the grain size observed in SEM images and the nominal thickness is currently under investigation.}
\label{CsISEM}
\end{figure*}

The results presented in this paper are obtained after applying several selection criteria (cuts) to the triggered events. 
Firstly, while the CFD method reduces time walk, a~residual dependence between SAT and the e-peak charge remains. This arises from avalanche fluctuations, which cause charge-dependent variations in SAT and degrade the time resolution.
The dependence is parameterised and used to correct the SAT by subtracting the fitted charge-dependent pedestal.
Moreover, a~time window cut selects events within 300~ps of the median time difference of all recorded signals, removing off-time events and noise fluctuations. 
Amplitude cuts are applied, rejecting events with signals below 1\% of the dynamic range, classified as empty, or above 99\%, categorised as saturated.
A geometrical cut is also applied due to the Cherenkov radiator emitting a~UV photon cone at approximately 45$^\circ$. 
To ensure accurate time resolution measurements, only events fully contained within the detector's active area are considered. 
For a~10~mm diameter detector with a~3~mm thick radiator, this corresponds to a~4~mm diameter circle around the pad center. 
Tracks outside this central region produce reduced signals due to partial photoelectron loss beyond the readout area. 
After these selections and corrections, the detection efficiency consistently exceeds 95\%.

%\subsubsection*{Number of photoelectrons}

The photocathode performance is further evaluated through the number of extracted photoelectrons, $N_{\text{PE}}$, defined as the ratio of the mean MIP signal charge to the mean SPE signal charge under identical detector settings.
SPE signals are recorded using a~light-emitting diode (LED) operated at low intensity to extract only one photoelectron at a~time, while MIP signals are measured with a~muon beam.
A smoothing filter is applied to the signals to suppress high-frequency fluctuations while preserving the peak structure~\cite{GuerraSummer}.
This simplified waveform enables reliable extraction of key features, including:
%\newline
(i)~more precise determination of the electron-peak start;
%\newline
(ii)~clear definition of the ion-tail end;
%\newline
(iii)~identification of multiple peaks, likely due to photon feedback;
%\newline
(iv)~rejection of noise events. %whose smoothed maximum amplitude is less than three times the baseline noise RMS. 
%\newline
Integrating the waveform from the electron-peak start to the ion-tail end yields charge distributions with a~clearer separation between noise and true SPE signals.
Each signal charge is extracted and used to create a~histogram, which is then fitted with a~Pólya distribution:
\begin{equation}
P_n~=~\frac{(\theta + 1)^{\theta + 1}}{\bar{n}\,\Gamma(\theta + 1)}
\left( \frac{n}{\bar{n}} \right)^{\theta}
e^{-(\theta + 1)n/\bar{n}},
\label{eq:Pólya}
\end{equation}
where $n$ is the number of charges produced, $\bar{n}$ the mean avalanche size and $\theta$ the shape parameter.
The mean charge for the MIP, $\bar{n}_{\text{MIP}}$, is divided by the mean charge for the SPE, $\bar{n}_{\text{SPE}}$, to obtain the number of photoelectrons:
\begin{equation}
N_{\text{PE}}~=~\frac{\bar{n}_{\text{MIP}}}{\bar{n}_{\text{SPE}}}.
\label{eq:NPE}
\end{equation}
Examples of SPE and MIP charge spectra, obtained with a 5 nm B$_4$C photocathode, are shown in Fig.~\ref{PolyaNPE}.

\section{Photocathode characterisation}
\label{sec:4}

The characterisation of four photocathode materials - CsI, Ti, B$_4$C and DLC - is presented.
The research included laboratory tests of optical and resistive properties, along with beam measurements to evaluate time resolution and photoelectron yield.
The studies involved scanning the cathode voltage while keeping the anode voltage fixed at \(V_A = 275~\mathrm{V}\) and the mesh at ground potential.
The highest cathode voltage used was always set to 10 V below the discharge threshold in the detector.
In each case, voltage scans were performed by decreasing the cathode voltage in 10~V steps from the maximum value, covering at least three voltage points.

To ensure uniform experimental conditions, the measurements were performed using an identical single-pad metallic detector featuring a~$\varnothing$10\,mm  active area and pre-amplification and amplification gaps of approximately 127\,$\mu$m, operated in sealed mode at a~gas pressure of 990 $\pm$ 5\,mbar.
The pressure of 990~mbar was chosen to keep conditions close to the flushing mode, corresponding to around 970 mbar, while maintaining a~slight overpressure to prevent ambient air from entering the detector, thereby preserving gas quality and ensuring stable operation.
As the Micromegas printed circuit board is mechanically decoupled from the outer enclosure and incorporates openings for gas circulation, it ensures equal pressure on both sides of the amplification structure. This configuration minimises mechanical deformation, thereby preserving the amplification gap and maintaining stable electric fields.
The uncertainty of $\pm$5 mbar was estimated, as the chamber is filled manually and slight under- or overshoots can occur.
A~detailed description of the mechanical design is provided in~\cite{UtrobicicSinglePad}.

For one CsI comparison measurements, a~prototype with a~surface resistivity of 20\,M$\Omega$/$\Box$ was used and operated in flushing  mode at an atmospheric pressure of approximately 970~mbar during the July test beam campaign. 
These measurements were initially performed to test the resistive detector and were subsequently also used to compare different CsI photocathode thicknesses.
The selected photocathode was then remeasured using a metallic prototype operated in sealed mode at a gas pressure of 990~$\pm$~5\,mbar during the November test beam campaign, to enable comparison with other photocathodes.

\begin{figure}[!t]
\begin{center}
\includegraphics[width=\columnwidth]{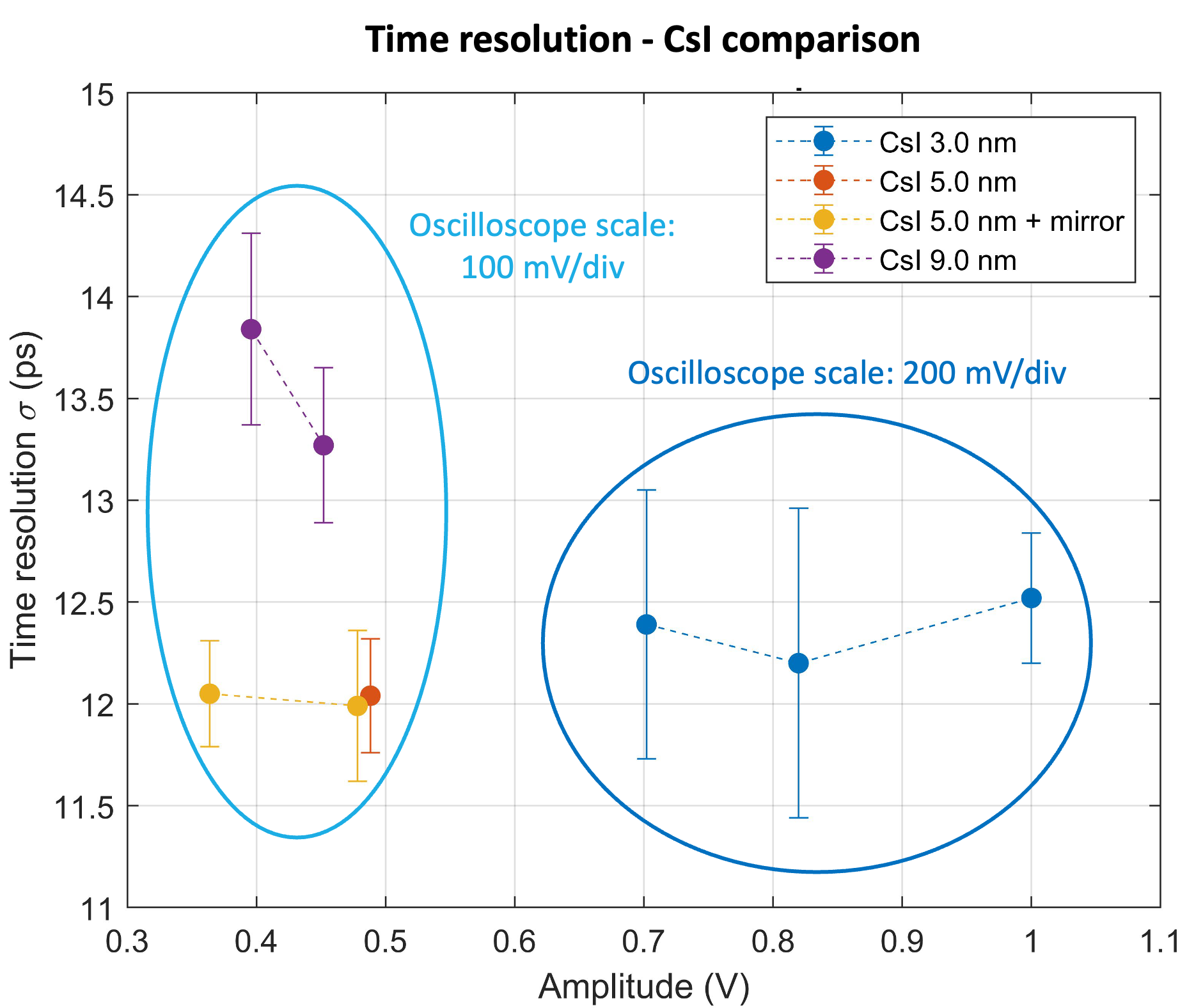}
\end{center}
\vspace{0.8mm}
\caption{Time resolution as a~function of mean signal amplitude for different CsI photocathode thicknesses. The samples were evaluated using a~resistive detector in flushing mode at 970 $\pm$ 5\,mbar. Thinner layers generally provided better time resolution. A~sample with a~mirror showed no improvement in time resolution compared to those without the reflective layer, indicating that internal reflections do not significantly contribute to the detected signal. For the 3 nm sample, very high signal amplitudes were observed, requiring an adjustment of the oscilloscope scale, which degraded the measured time resolution.}
\label{CsIComparison}
\end{figure}

\subsection{Cesium Iodide}

CsI is commonly used as a~photocathode material in gaseous detectors due to its high QE and strong sensitivity to UV radiation.
It serves as a~UV-to-electron converter in major RICH systems, including those in ALICE  \cite{ALICE_RICH, ALICE_CsI} and COMPASS~\cite{COMPASS_RICH_CsI}.
The first PICOSEC Micromegas detector employed an 18 nm CsI layer deposited on a~3 mm MgF$_2$ substrate with a~3 nm Cr interlayer, demonstrating a~time resolution of $\sigma$~<~25 ps~\cite{firstPicosecPaper}.

Although the CsI photocathode is not a~robust long-term solution, it remains valuable for studying the detector and electronics performance.
Extensive studies of CsI deposition and storage were carried out at the TFG workshop to ensure its high-quality performance.
These efforts established an optimised preparation procedure based on cleaning the substrate with high-purity solvents: acetone and 99.9\% ethanol.
After deposition, the CsI samples were stored under vacuum or in a~dry gas environment to preserve their timing performance, as exposure to humidity degrade the photocathode within minutes.

Previous studies of VUV transparency have shown that bare 3 nm MgF$_2$ radiator transmits approximately 80\% at 160 nm, which decreases to 45\% with a~3\,nm Cr layer and drops further to 10\% after deposition of the 18\,nm CsI film~\cite{myMScThesis}.
Thinner CsI layers - 9\,nm, 5\,nm and 3\,nm - were recently investigated, all deposited on 3\,mm MgF$_2$ substrates with a 2.4~nm Ti interfacial layer in between, which serves as a HV contact and mitigates charging-up effects.
SEM measurements were performed to compare the microscopic morphology of CsI films of different thicknesses, as illustrated in Fig.~\ref{CsISEM}. All samples were briefly exposed to air during detector assembly or transport to the SEM, for a~total of no more than 2~minutes. The images show a~grain-like structure, with grain size increasing with nominal photocathode thickness. This agglomeration reflects the hygroscopic nature of CsI and its sensitivity to humidity. 
Consequently, the thinnest layer exhibits the most uniform structure. The mismatch between the grain size observed in SEM images and the nominal thickness is currently under investigation.
It should be noted that these CsI layers were significantly thinner than those used in large-area detectors such as the ALICE~\cite{ALICE_RICH, ALICE_CsI} and COMPASS~\cite{COMPASS_RICH_CsI} RICH systems, where CsI photocathodes with thicknesses on the order of hundreds of nanometres were operated in reflective mode. SEM measurements performed on CsI films that were fully protected from air exposure showed that film growth on the substrate proceeded from well-separated nucleation spots to a~homogeneous coverage once the thickness exceeded approximately 100~nm~\cite{ALICE_RICH}.
In contrast, the semi-transparent CsI layers used in PICOSEC Micromegas detectors are only a~few nanometres thick, leading to different growth conditions and surface morphologies. Therefore, a~direct comparison of granularity and uniformity between these systems was not straightforward.

\begin{figure}[!t]
\begin{center}
\includegraphics[width=\columnwidth]{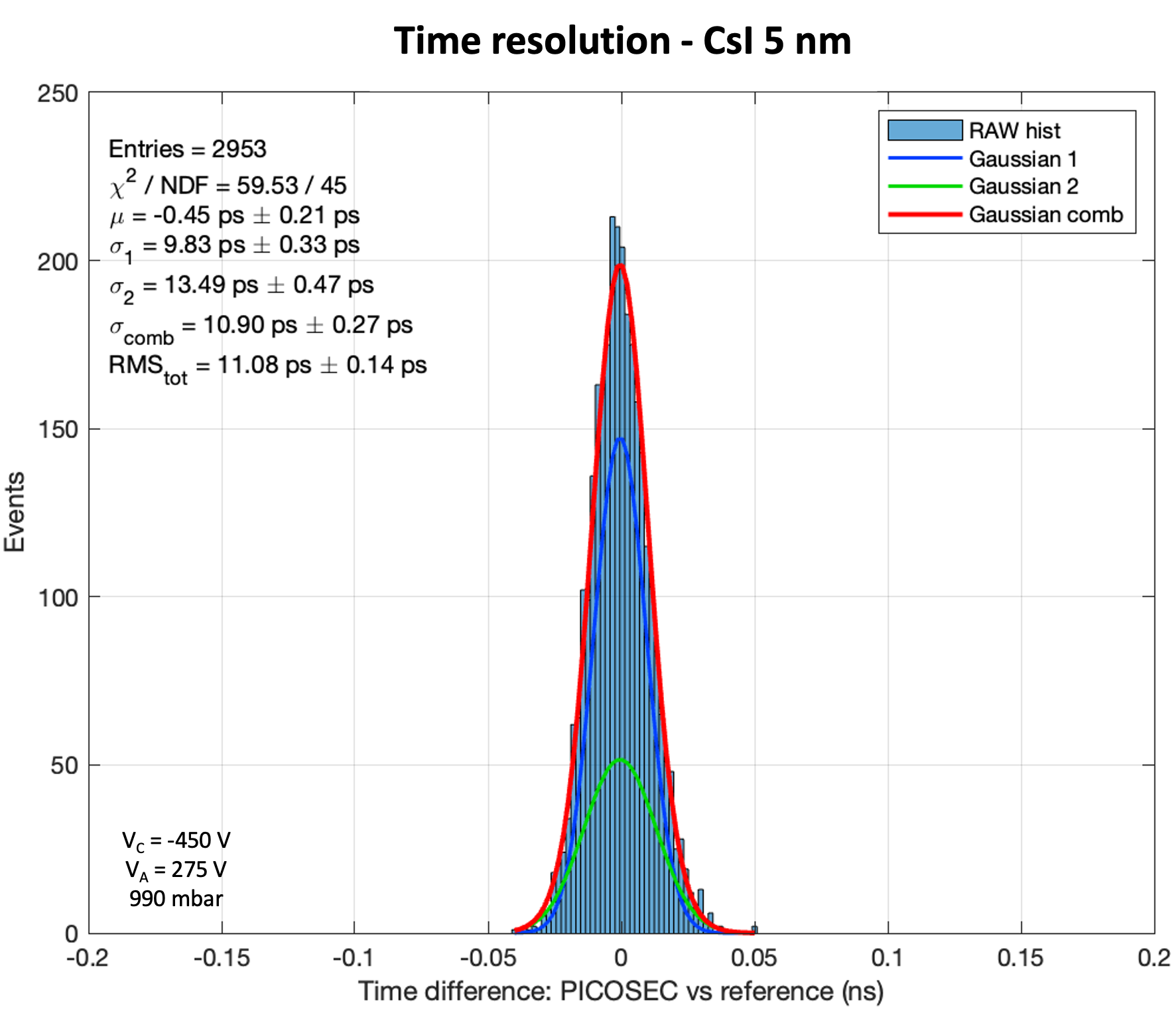}
\end{center}
\caption{
SAT distribution of the single-pad metallic detector with a~10\,mm active area, equipped with a~5\,nm CsI photocathode with a~2.4\,nm Ti  interlayer, operated in sealed mode at 990 $\pm$ 5\,mbar. The applied voltages were V$_\text{C}$~=~$-450$\,V and V$_\text{A}$~=~275\,V. The histogram includes only fully contained events selected by a~4\,mm diameter geometrical cut around the channel centre. The time resolution is $\sigma~=~10.9 \pm 0.3$\,ps with a~99.9\% detection efficiency.
}
\label{CsITimeRes}
\end{figure}

\begin{figure*}[!t]
\begin{center}
\includegraphics[width=\textwidth]{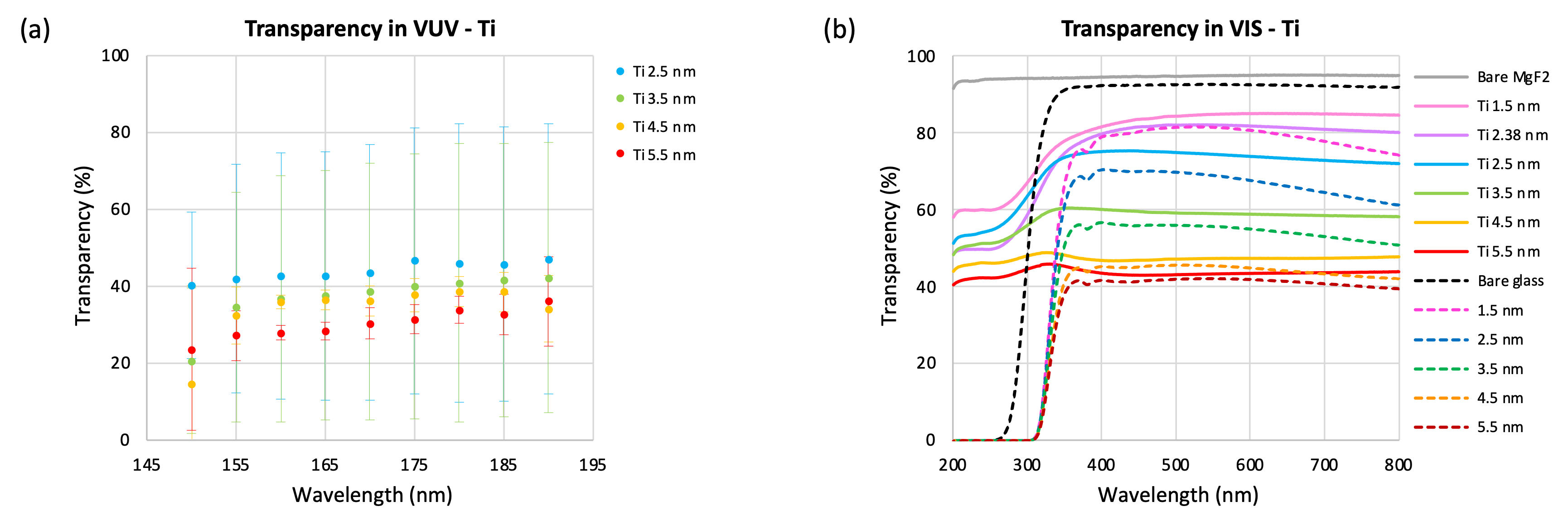}
\end{center}
\caption{Transparency as a~function of wavelength in (a) the VUV and (b) the VIS ranges of Ti photocathodes deposited on two different substrates, MgF$_2$ and glass. A~strong thickness dependence is observed, with the transparency at 160\,nm increasing from about 28\% for the thickest film to 43\% for the thinnest.}
\label{TiTransparency}
\end{figure*}

\begin{figure}[!t]
\begin{center}
\includegraphics[width=\columnwidth]{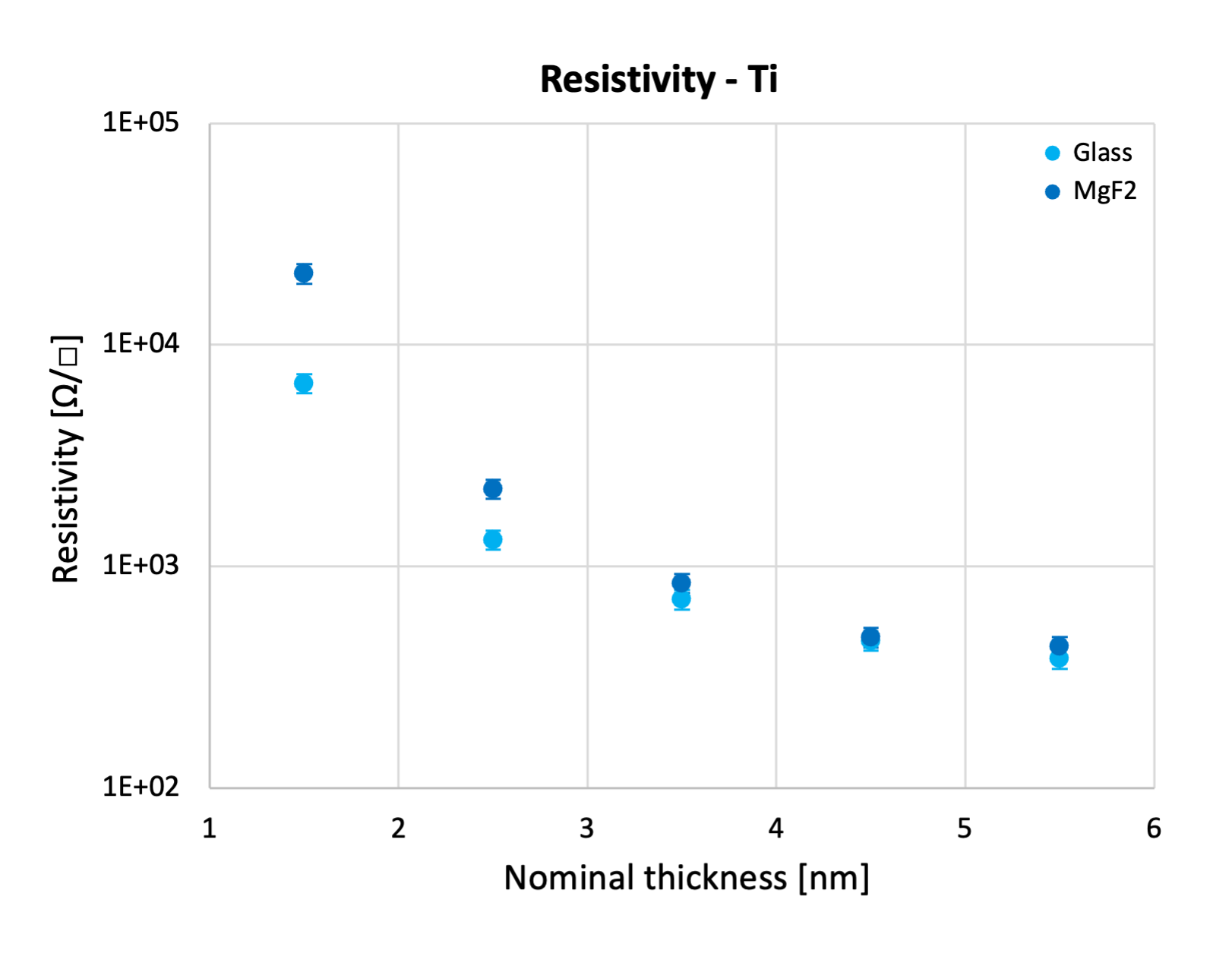}
\end{center}
\caption{Surface resistivity versus nominal thickness for Ti layers deposited on MgF$_2$ and glass, varying from ~400\,$\Omega/\square$ for the thickest films to ~20\,k$\Omega/\square$ for the thinnest. Higher resistivity on MgF$_2$ compared to glass suggests thinner layers, possibly due to lower crystal adhesion.}
\label{TiResistivity}
\end{figure}

The samples were initially evaluated using a~resistive detector, with measurements performed in flushing mode at an atmospheric pressure of 970 $\pm$ 5\,mbar in the July test beam campaign.
%The cathode voltage was scanned while the anode voltage was kept fixed at \(V_A=275~\mathrm{V}\).
The comparison across different coating thicknesses showed that thinner layers generally provide better time resolution, as presented in Fig.~\ref{CsIComparison}.
One sample featured a~100~nm aluminium mirror layer deposited on the MgF$_2$ side opposite to the photocathode. Cherenkov photons produced in the radiator and directed towards the photocathode can either be transmitted through it or reflected at its surface. In this configuration, the reflected photons can reach the opposite side of the crystal, where they are reflected again by the mirror and then propagate back towards the photocathode. This approach was intended to increase the effective photon yield and improve the timing performance. However, no improvement in time resolution was observed compared to samples without the reflective layer, indicating that such multiple reflections within the radiator do not significantly contribute to the detected signal.
For the 3~nm sample, very high signal amplitudes were observed, requiring an adjustment of the oscilloscope scale to prevent signal saturation, which resulted in a degradation of the measured time resolution due to coarser sampling at larger oscilloscope scales. This demonstrates that digitiser settings can significantly influence the final timing performance.

The best result was achieved with the 5\,nm CsI photocathode, reaching a~time resolution of $\sigma = 12.0 \pm 0.3$\,ps at a~cathode voltage of \(V_\text{C} = -455~\mathrm{V}\). The photocathode was remeasured using a~metallic prototype operated at \(V_\text{C} = -450~\mathrm{V}\) in sealed mode at a~gas pressure of 990 $\pm$ 5\,mbar in the November test beam campaign.
Under these conditions, a~time resolution of $\sigma~=~10.9~\pm~0.3$\,ps was obtained with a~99.9\% detection efficiency for fully contained events, as shown in Fig.~\ref{CsITimeRes}. This represents the most precise timing result achieved by PICOSEC Micromegas to date.
The number of photoelectrons extracted from the 5\,nm CsI layer was measured as $N_\text{PE} = 32.35 \pm 0.35$ per MIP.

\subsection{Titanium}

Ti was studied both as a~stand-alone photocathode material and as an interlayer for CsI. Its higher work function and roughly 15\% greater transparency compared to Cr make it an attractive option, as the increased transparency allows more photons to reach the photocathode. Additionally, Ti films are insensitive to humidity and therefore do not require controlled storage conditions.
Encouraging results from initial Ti tests~\cite{LisowskaPhDThesis} prompted more detailed follow-up studies.

Ti samples were produced at the TFG workshop. Films with nominal thicknesses between 1.5\,nm and 5.5\,nm were deposited on 3\,mm MgF$_2$ and glass substrates for comparison. Transparency measurements, shown in Fig.~\ref{TiTransparency}, exhibited a~strong thickness dependence, increasing from about 28\% at 160\,nm for the thickest film to 43\% for the thinnest. Surface resistivity, presented in Fig.~\ref{TiResistivity}, ranged from approximately 400\,$\Omega/\square$ for the thickest layers to about 20\,k$\Omega/\square$ for the thinnest.
Higher resistivity values for photocathodes deposited on MgF$_2$ compared to glass indicate a reduced effective thickness, likely due to poorer adhesion of Ti on MgF$_2$. As a result, a fraction of the deposited material does not adhere to the substrate and is instead lost to the chamber walls during deposition.

 \begin{figure}[!t]
\begin{center}
\includegraphics[width=\columnwidth]{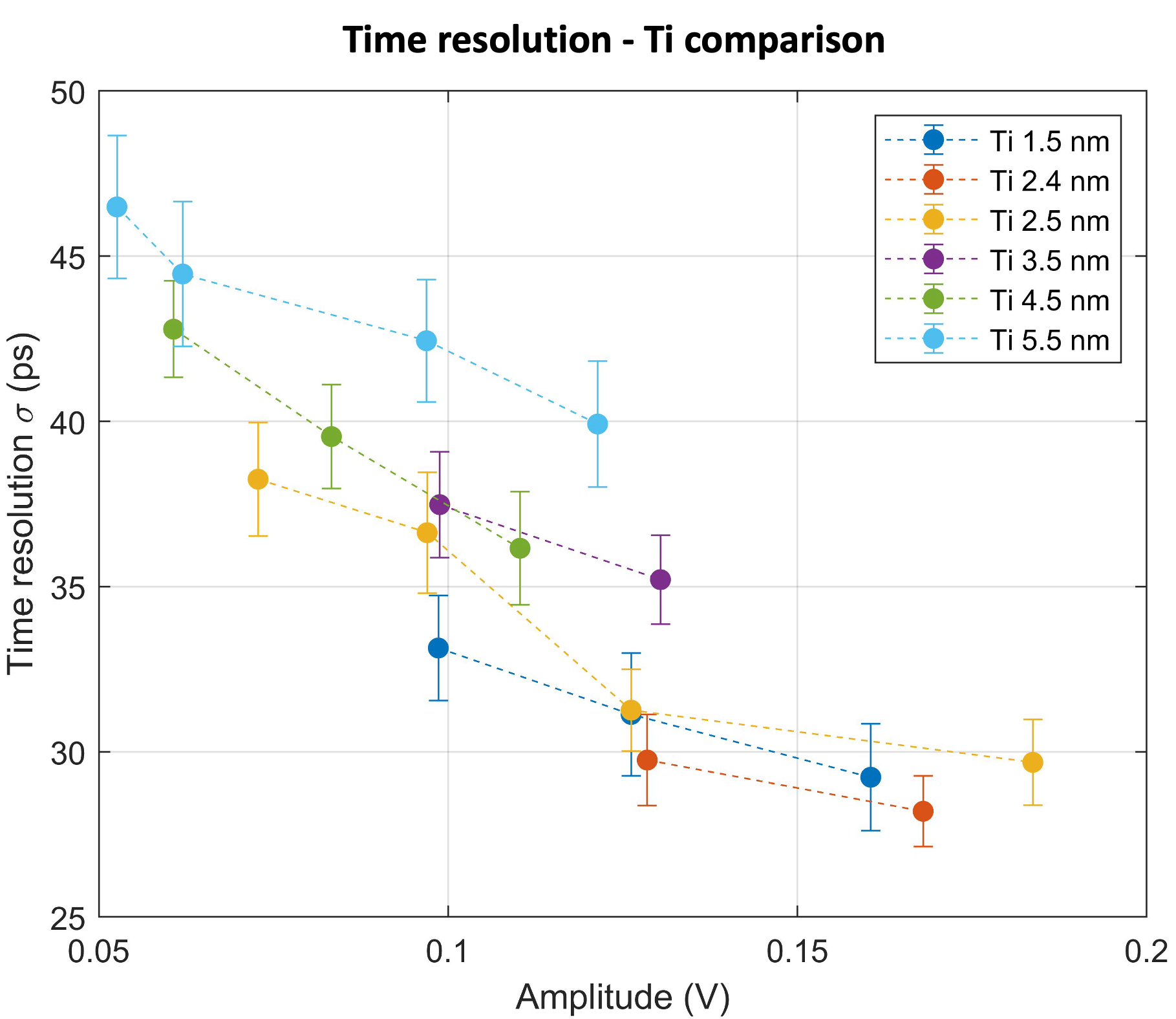}
\end{center}
\caption{Time resolution as a~function of mean signal amplitude for different Ti photocathode thicknesses. The samples were evaluated with a~metallic detector in sealed mode at a~gas pressure of 990 \(\pm\) 5\,mbar, showing that thinner layers generally yield better time resolution.}
\label{TiComparison}
\end{figure}

\begin{figure}[!t]
\begin{center}
\includegraphics[width=\columnwidth]{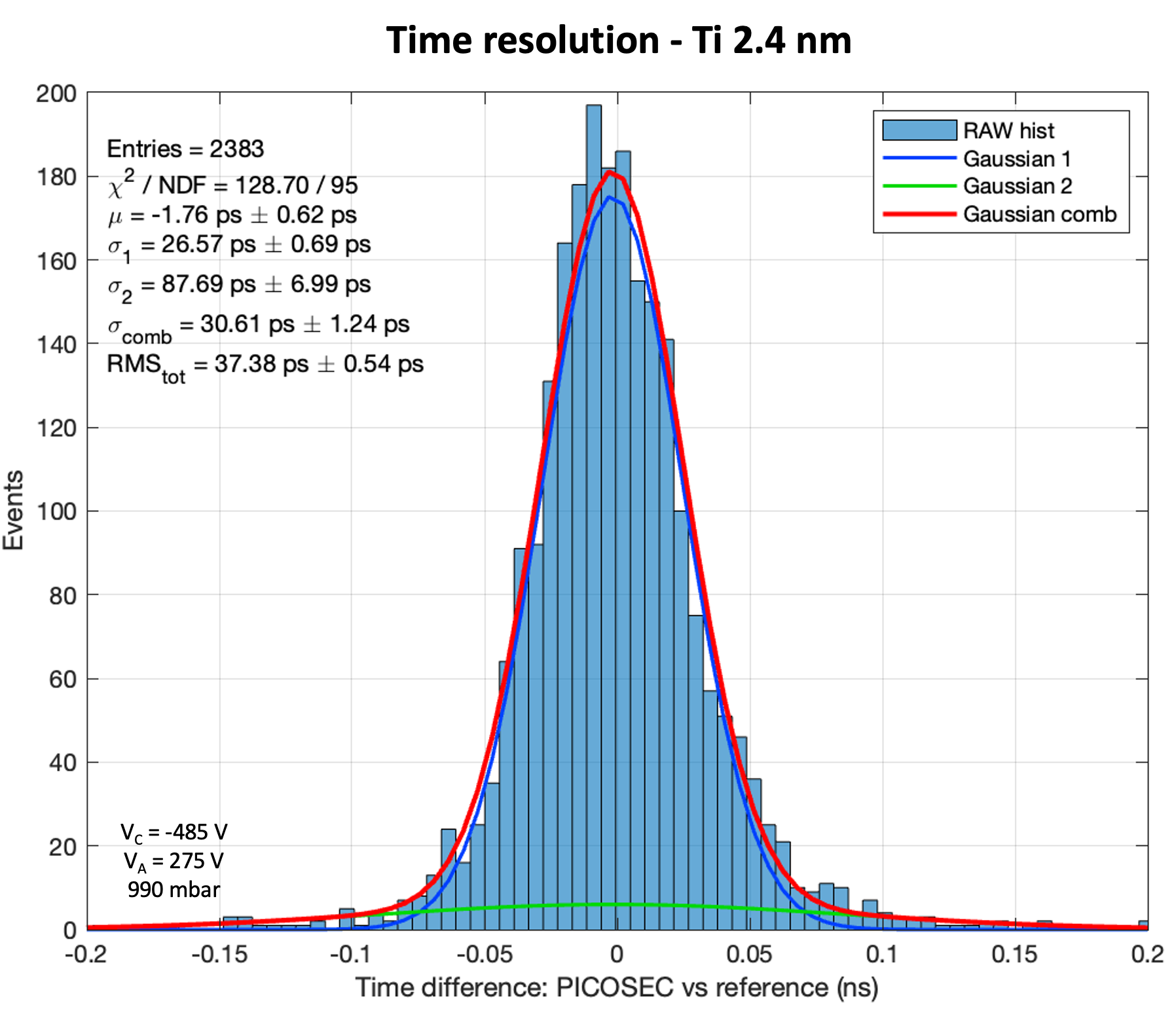}
\end{center}
\vspace{0.5mm}
\caption{SAT distribution of the metallic prototype assembled with a~2.4\,nm Ti photocathode, operated in sealed mode at 990 $\pm$ 5\,mbar. The applied voltages were \(V_\text{C} = -485\)\,V and \(V_\text{A} = 275\)\,V. The time resolution is \(\sigma~=~30.6~\pm~1.2\)\,ps with a~98.8\% detection efficiency.}
\label{TimeResTi}
\end{figure}

The samples were evaluated with the same metallic detector described above, operated in sealed mode at a~gas pressure of 990 $\pm$ 5\,mbar during the July test beam campaign.
Scanning the cathode voltage for different photocathode thicknesses again showed that thinner layers generally yield better time resolution, as illustrated in Fig.~\ref{TiComparison}.
The best-performing photocathode, a~2.4\,nm Ti layer, was remeasured during the November campaign.
A~time resolution of $\sigma = 30.6 \pm 1.2$\,ps was achieved at a~cathode voltage of \(V_\text{C} = -485~\mathrm{V}\), with a~98.8\% detection efficiency, as presented in Fig.~\ref{TimeResTi}.
Measurements indicated that the 2.4\,nm Ti film produced $N_\text{PE} = 5.10 \pm 0.05$ photoelectrons per MIP.

\subsection{Boron Carbide}

B$_4$C was investigated as a~more robust alternative to CsI for photocathode applications.
Initial B$_4$C photocathodes were developed at the French Atomic Energy Commission (CEA) and the European Spallation Source (ESS), with various thicknesses tested, yielding promising results~\cite{LukasPhDThesis,myMScThesis,LisowskaPhDThesis,myPhotocathodes}. These findings motivated further studies, which are presented in this paper.

\begin{figure*}[!t]
\begin{center}
\includegraphics[width=\textwidth]{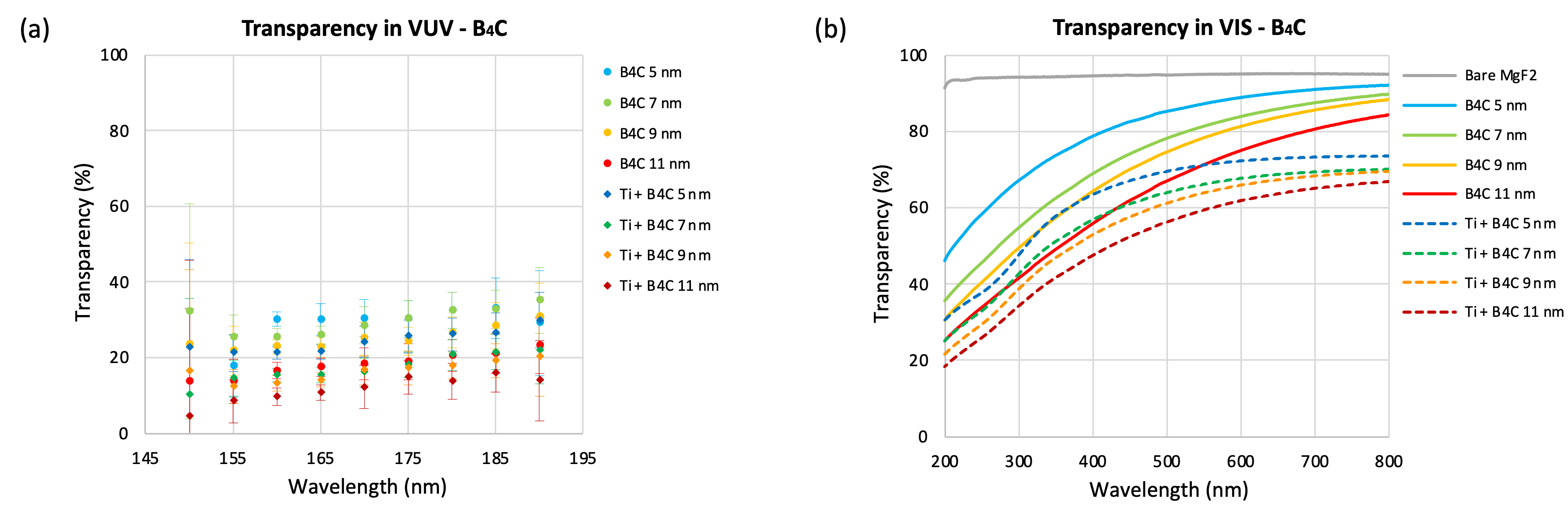}
\end{center}
\caption{
Transparency versus wavelength in the VUV (a) and VIS (b) ranges for B$_4$C photocathodes on MgF$_2$ substrates. Samples without a~Ti interlayer show transparencies of about 15\% at 160\,nm for the thickest films, increasing to 30\% for the thinnest. The addition of the Ti interlayer reduces transparency by 30\%.} 
\label{B4CTransparency}
\end{figure*}

\begin{figure}[!t]
\begin{center}
\end{center}
\vspace{0.6mm}
\includegraphics[width=\columnwidth]{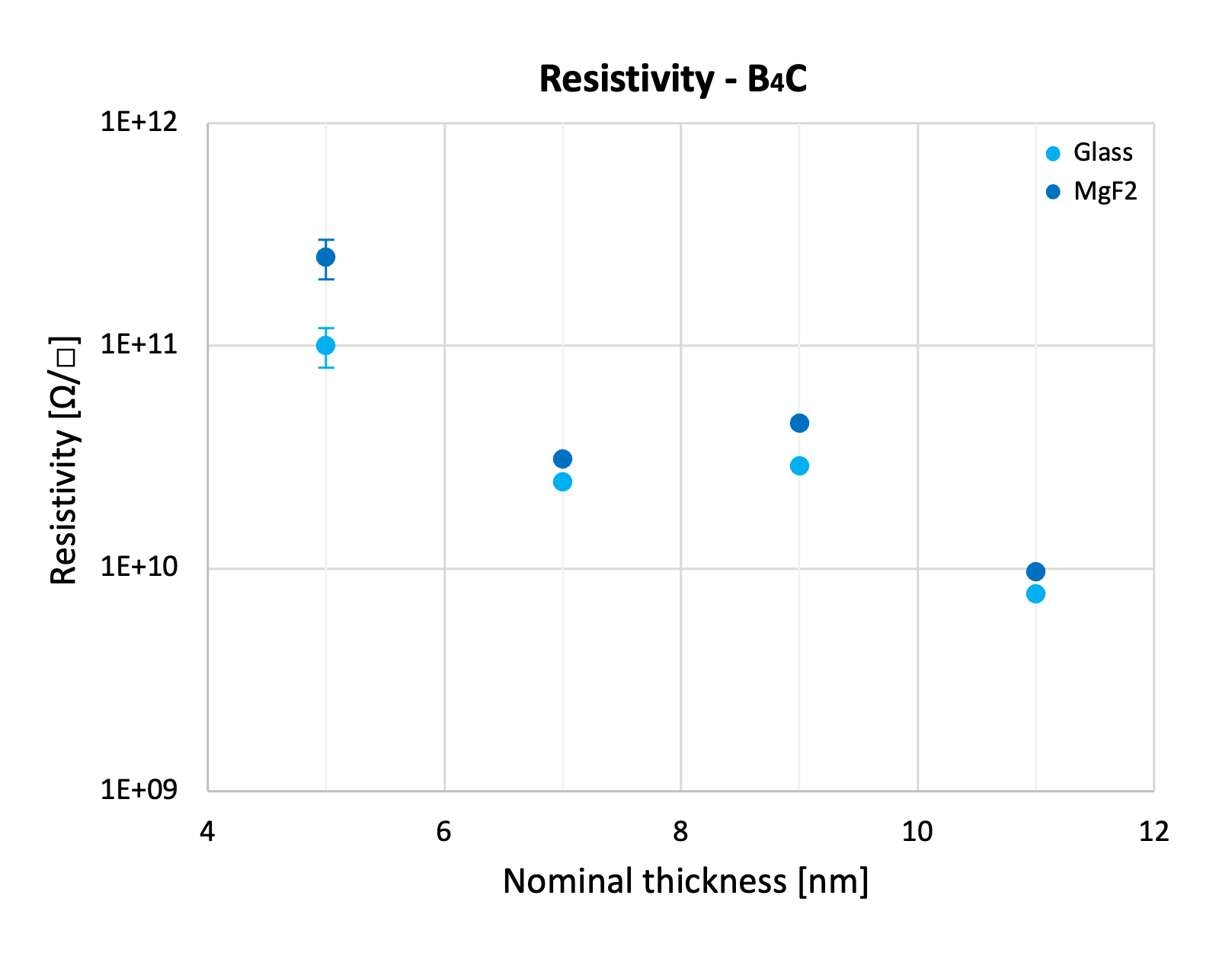}
\vspace{1mm}
\caption{Surface resistivity as a~function of nominal thickness for B$_4$C photocathodes without Ti layers, showing very high values in the range of 10–100~G$\Omega/\square$. The deviations from the expected trend may arise from non-uniformity of the thin layers.}
\label{B4CResistivity}
\end{figure}

The samples described in this paper  were fabricated by magnetron sputtering at the MPT workshop.
%Since this system does not allow direct measurement of the deposited film thickness, thin layers (a few nanometres) were produced by scaling the deposition time based on calibration runs with thicker films (tens of nanometres).
The photocathodes with nominal thicknesses between 5\,nm and 11\,nm were deposited  on 3\,mm MgF$_2$ substrates, with some incorporating a~2.4\,nm Ti interfacial layer. 
Transparency measurements of B$_4$C photocathodes again revealed a~clear thickness dependence: samples without an interlayer showed transparencies of about 15\% at 160\,nm for the thickest films and up to 30\% for the thinnest, as shown in Fig.~\ref{B4CTransparency}.
The addition of a~Ti layer reduced the transparency by approximately 30\%.
Surface resistivity measurements of the B$_4$C photocathodes without Ti layer indicated very high values, in the range of 10-100\,G$\Omega/\square$, as presented in Fig.~\ref{B4CResistivity}.
The deviations from the expected trend may be attributed to the potential non-uniformity of these thin layers.

\begin{figure}[!t]
\begin{center}
\includegraphics[width=\columnwidth]{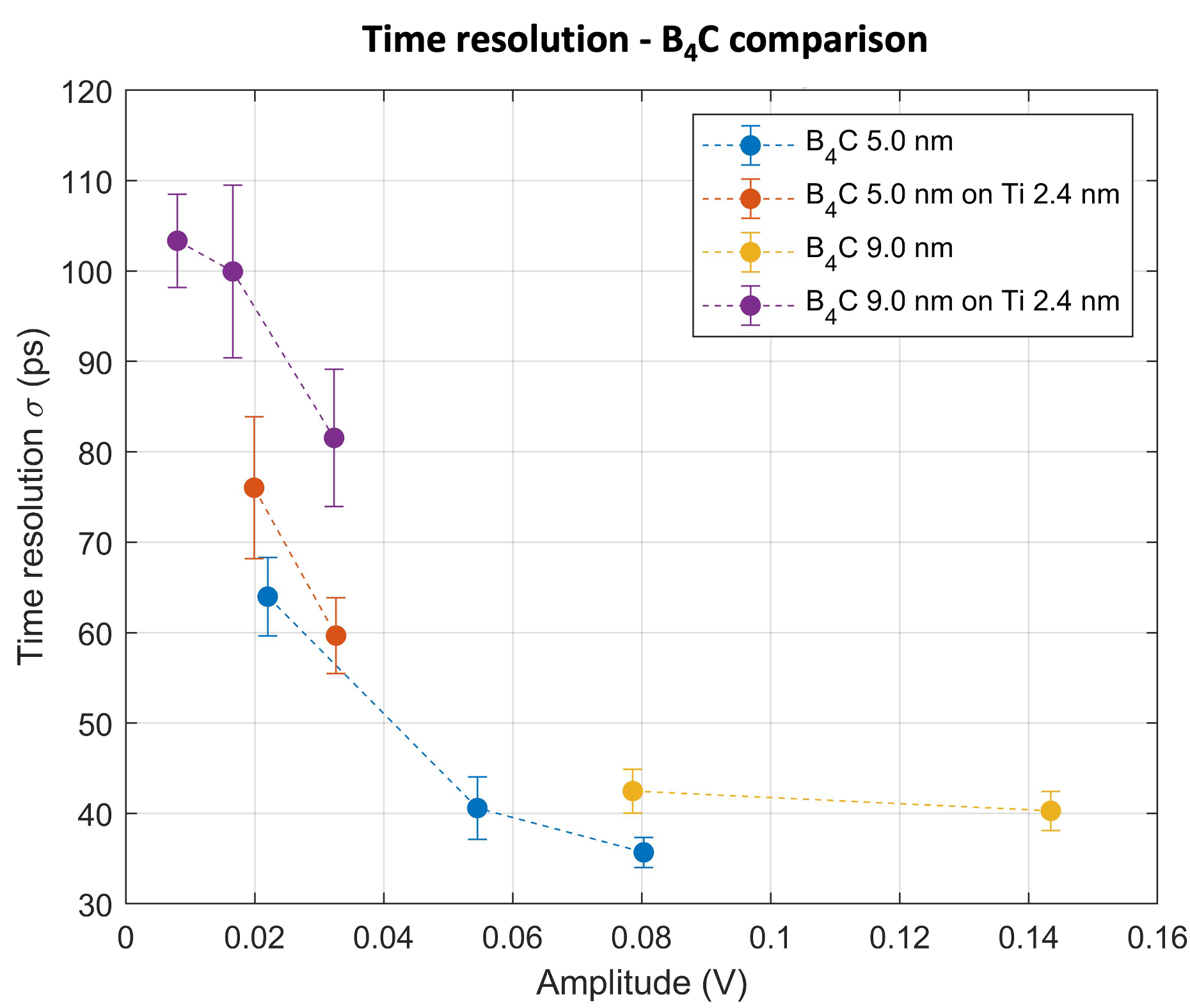}
\end{center}
\caption{Time resolution versus mean signal amplitude for different B$_4$C photocathode thicknesses, measured with the metallic detector in sealed mode at 990 $\pm$ 5\,mbar, showing improved performance for photocathodes without Ti layers. The degradation in timing resolution for photocathodes with Ti layers can be attributed to their reduced transparency, leading to fewer detected photoelectrons.}
\label{B4CComparison}
\end{figure}

\begin{figure}[!t]
\begin{center}
\includegraphics[width=\columnwidth]{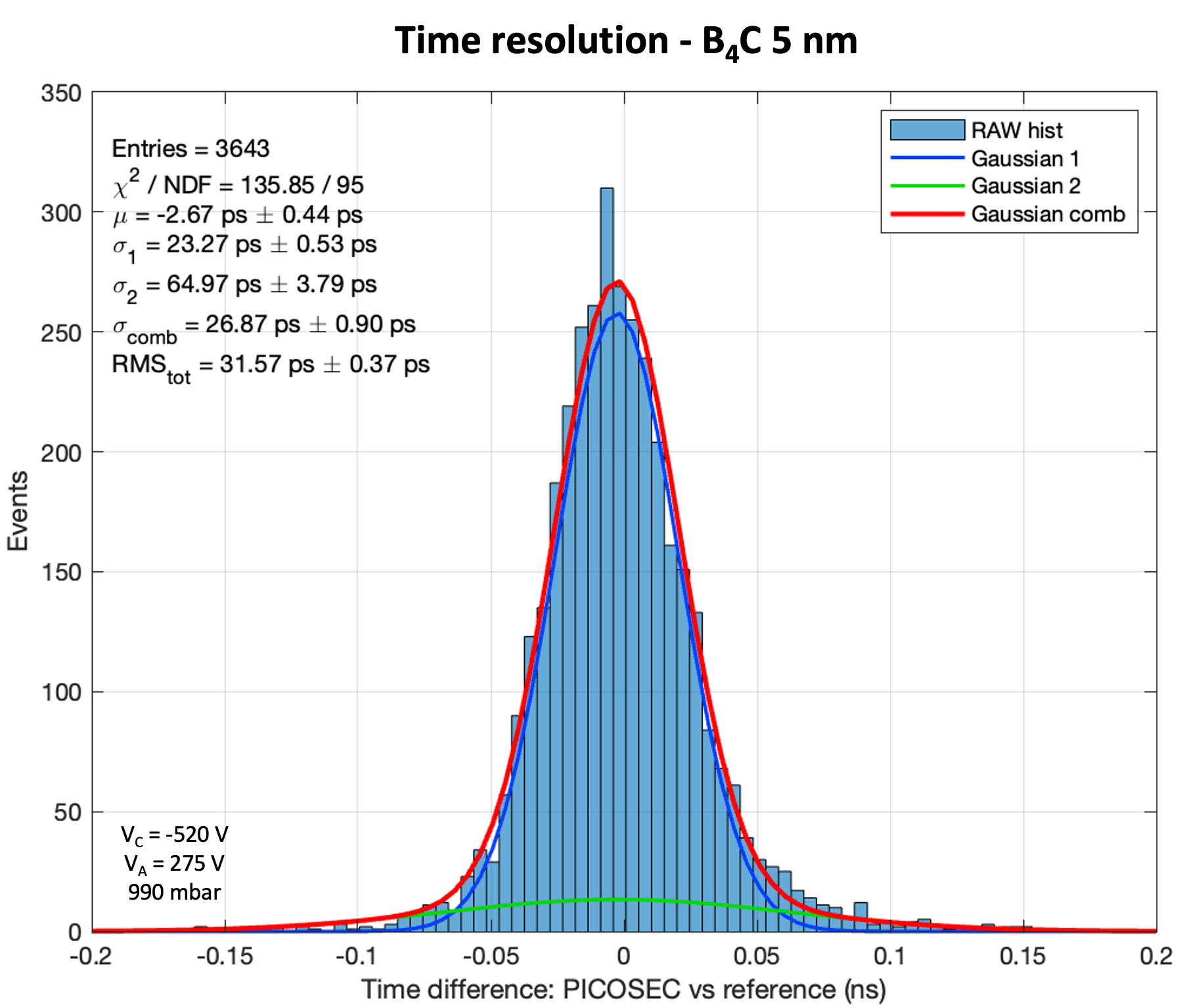}
\end{center}
\caption{SAT distribution of the detector equipped with a~5\,nm B$_4$C photocathode. The applied voltages were V$_\text{C}$~=~$-520$\,V and V$_\text{A}$~=~275\,V and the device was operated in the sealed mode at 990 $\pm$ 5\,mbar. The time resolution is $\sigma~=~26.9 \pm 0.9$\,ps  with a~99.1\% detection efficiency.}
\label{TimeResB4C}
\end{figure}

The samples were evaluated using the metallic prototype operated in sealed mode during the July test beam campaign.
A scan of the cathode voltage for different film thicknesses showed that photocathodes without a~Ti interlayer provide better time resolution, as illustrated in Fig.~\ref{B4CComparison}.
The observed degradation in time resolution for photocathodes with Ti layers can be attributed to their reduced transparency, leading to a~lower number of detected photoelectrons.

The best performance was obtained with the 5\,nm B$_4$C photocathode, reaching a~time resolution of $\sigma = 35.7 \pm 1.7$\,ps at \(V_\text{C} = -490~\mathrm{V}\).
After storage in air, the photocathode was remeasured during the November campaign under identical detector conditions to verify the reproducibility of earlier results and to determine the photoelectron yield.
The signal amplitude increased and the time resolution improved to $\sigma = 29.0 \pm 1.2$\,ps, possibly due to oxidation of the layer.
At the highest stable operating voltage of \(V_\text{C} = -520~\mathrm{V}\), the prototype achieved $\sigma = 26.9 \pm 0.9$\,ps with a~99.1\% detection efficiency, as shown in Fig.~\ref{TimeResB4C}. The number of photoelectrons extracted from the 5\,nm B$_4$C layer was measured as $N_\text{PE} = 5.43 \pm 0.04$ per MIP.
Repeated measurements confirmed the reproducibility of these results. 
The effect of oxidation on B$_4$C is currently under investigation, in particular by introducing controlled amounts of oxygen during the B$_4$C sputtering deposition to determine whether this mechanism is responsible for the observed improvement.

%\begin{figure}[!b]
%\begin{center}
%\includegraphics[width=\columnwidth]{figures/TimingB4C5nm.png}
%\end{center}
%\caption{SAT distribution of the prototype equipped with a~5\,nm B$_4$C photocathode. The applied voltages were V$_\text{C}$~=~$-490$\,V and V$_\text{A}$~=~275\,V. The measured time resolution is $\sigma~=~35.7 \pm 1.7$\,ps.}
%\label{TimeResB4C}
%\end{figure}

\subsection{Diamond-Like Carbon}

Initial DLC photocathodes were developed and characterised at the University of Science and Technology of China (USTC) and at CERN, consistently demonstrating strong performance and robustness, making them excellent candidates for PICOSEC photocathodes~\cite{LukasPhDThesis,myMScThesis,XuPhotocathodes,LisowskaPhDThesis,myPhotocathodes}.
The results reported in~\cite{LisowskaPhDThesis,myPhotocathodes} showed that films with nominal thicknesses of a~few nanometers fabricated by magnetron sputtering at the MPT workshop exhibited approximately 60\% transparency at 160\,nm and very high surface resistivity values ranging from 1 to 100\,G$\Omega/\square$. Time resolution measurements indicated that thinner layers provided better performance, with the 1.5\,nm DLC photocathode reaching $\sigma = 31.9 \pm 1.3$\,ps at \(V_\text{C} = -500~\mathrm{V}\) and 96.8\% detection efficiency.
DLC photocathodes with a~2.4~nm metallic (Cr instead of Ti) interlayer, measured in the same previous study, showed a~reduction in transparency of approximately 30\% and a degradation in timing resolution of about 2~ps.
These findings provided the basis for the new measurements presented below.

\begin{figure}[!t]
\begin{center}
\includegraphics[width=\columnwidth]{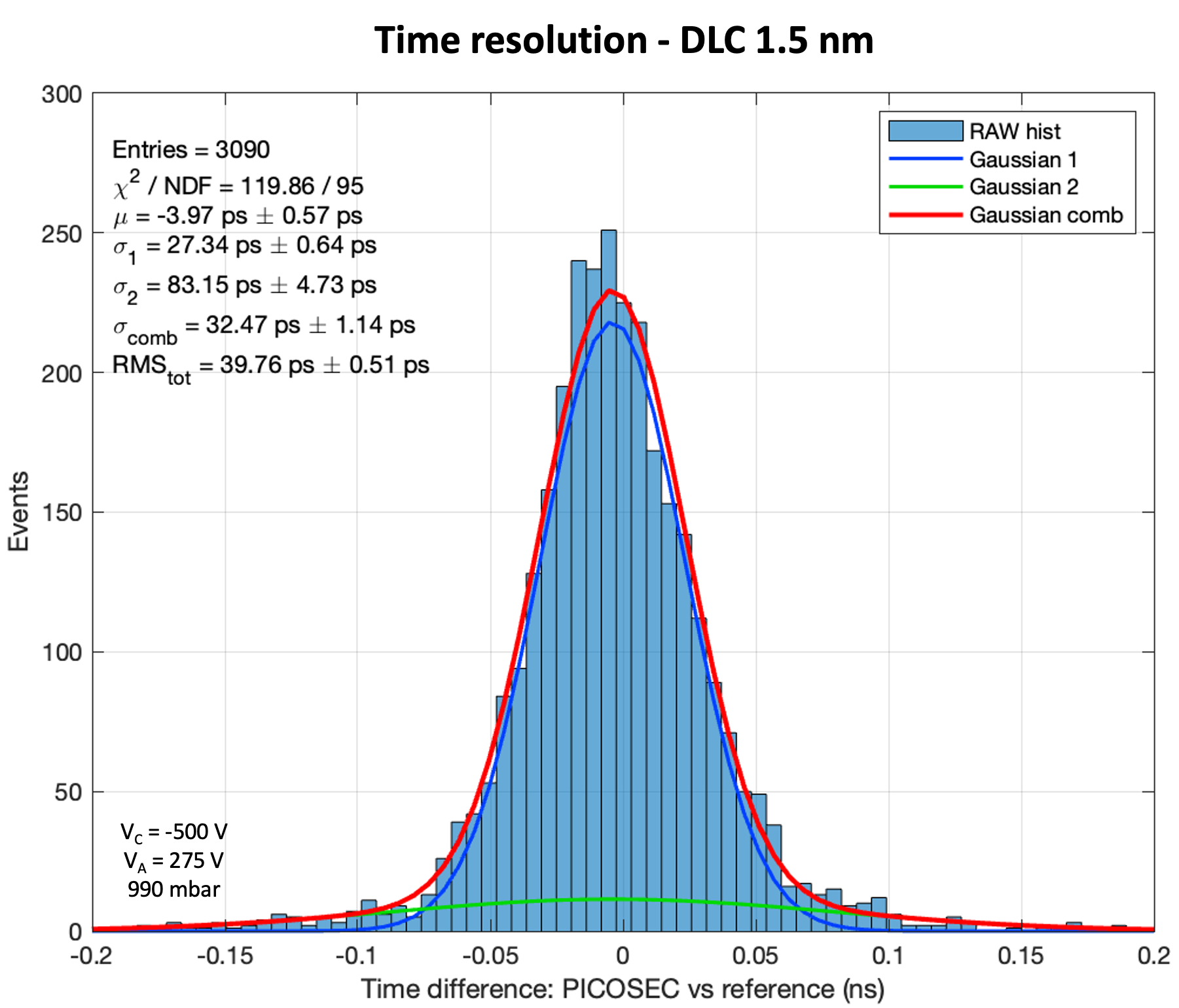}
\end{center}
\caption{SAT distribution of the prototype assembled with a~1.5\,nm DLC photocathode. The applied voltages were V$_\text{C}$~=~$-500$\,V and V$_\text{A}$~=~275\,V and the detector was operated in the sealed mode at 990 $\pm$ 5\,mbar. The time resolution is $\sigma~=~32.5 \pm 1.1$\,ps with a~96.8\% detection efficiency.}
\label{TimeResDLC}
\end{figure}

The 1.5~nm DLC photocathode was remeasured with the metallic detector during the November test beam campaign to assess the reproducibility of earlier results and to determine the photoelectron yield.
At the highest stable operating voltage of \(V_\text{C} = -500~\mathrm{V}\), the prototype achieved a~time resolution of $\sigma~=~32.5~\pm~1.1$~ps with a~96.9\% detection efficiency, as shown in Fig.~\ref{TimeResDLC}. Measurements show that the 1.5\,nm DLC layer yielded $N_\text{PE}~=~3.73~\pm~0.04$ photoelectrons per MIP.
The influence of the metallic interlayer on rate capability is not yet quantified and comparative studies with and without Ti are planned.

\section{Discussion}
\label{sec:5}

In the comparative analysis of photocathode materials, distinct performance characteristics were observed among CsI, Ti, B$_4$C and DLC, as summarised in Table~\ref{tab:Summary}.
Measurements on the muon beams indicated that thinner coatings generally provided superior timing performance.
CsI achieved the best time resolution of $\sigma = 10.9 \pm 0.3$\,ps. % representing the most precise timing result reported to date.
Among the alternative materials, B$_4$C performed best with $\sigma \approx 27$\,ps, followed by Ti with $\sigma \approx 30.5$\,ps and DLC with $\sigma \approx 32.5$\,ps.
Detection efficiency for fully contained events consistently exceeded 95\% for all materials.
Finally, CsI exhibited the highest $N_\text{PE}$, exceeding 30 photoelectrons per MIP, while Ti and B$_4$C showed comparable values above 5 and DLC exhibited values above 3.

\begin{table*}[!t]
\centering
\caption{Summary of the results obtained for different photocathodes: CsI, Ti, B$_4$C and DLC. To ensure uniform experimental conditions, all measurements were performed using the same single-pad metallic detector with a~$\varnothing$10\,mm active area, operated in sealed mode at a~gas pressure of 990 $\pm$ 5\,mbar.}
\begin{tabular}{|l|c|c|c|c|c|c|}
\hline
Photocathode & Anode & Cathode &  Time resolution $\sigma$ & Time resolution RMS & Efficiency & Average N$_\text{PE}$ \\
\hline
CsI 5.0 nm on Ti 2.4 nm  & 275 V & 445 V &  $10.9 \pm 0.3$\,ps & $11.1 \pm 0.1$\,ps & 99.9 \% & $32.35 \pm 0.35$ \\
Ti 2.4 nm  & 275 V & 485 V  &  $30.6 \pm 1.2$\,ps & $37.4 \pm 0.5$\,ps  & 98.8 \% & $5.10 \pm 0.05$  \\
B$_4$C 5.0 nm & 275 V & 520 V &  $26.9 \pm 0.9$\,ps & $31.6 \pm 0.4$\,ps & 99.1 \% & $5.43 \pm 0.04$  \\
DLC  1.5 nm & 275 V & 500 V & $32.5 \pm 1.1$\,ps & $39.8 \pm 0.5$\,ps  & 96.9 \% & $3.73 \pm 0.04$  \\
\hline
\end{tabular}
\label{tab:Summary}
\end{table*}

Although CsI demonstrated excellent time resolution, its QE degrades rapidly due to ion backflow and discharges.
Ti and B$_4$C currently represent the most promising alternatives.
Ti maintains excellent, reproducible time resolution, is easy to produce and is conductive, making it suitable for high-rate environments.
It is also insensitive to humidity and can be stored in air without visible or performance degradation.
In the case of B$_4$C, signal amplitude increased and time resolution improved after air exposure, likely related to surface oxidation; the effect will be investigated in future studies.
Both metallic and carbon-based samples have been reused across multiple test beam campaigns over several years, showing no deterioration in time resolution.

\section{Conclusions}
\label{sec:6}

This work presents advances in photocathode development for PICOSEC Micromegas precise-timing detectors.
The study characterises not only CsI but also metallic and carbon-based materials, including Ti, B$_4$C and DLC.
It combines laboratory measurements of optical transparency and surface resistivity with beam tests using a~single-pad metallic detector equipped with photocathodes and exposed to 150\,GeV/$c$ muons to evaluate time resolution and photoelectron yield.

The best performance was achieved with a~5\,nm CsI photocathode, reaching $\sigma = 10.9 \pm 0.3$\,ps with more than 30 extracted photoelectrons, representing the most precise timing result obtained with PICOSEC Micromegas to date.
Alternative materials also showed excellent performance, with Ti and B$_4$C emerging as the most promising candidates, achieving $\sigma \approx 30$\,ps with about 5 photoelectrons.
The reported time resolutions include all contributions from Equation \ref{eq:timeResolution}, in particular those from the reference detector and the readout electronics. This implies that the intrinsic time resolution of the PICOSEC Micromegas detector, for prototypes assembled with CsI photocathodes, is below 10 ps. At this stage, rather than pushing further below the 10 ps level, the emphasis is on improving the detector robustness.
Metallic and carbon-based photocathodes exhibit significantly greater resistance to ion backflow, discharges and humidity.
Overall, these results demonstrate that enhanced robustness can be achieved while maintaining excellent timing performance, supporting the feasibility of the PICOSEC Micromegas concept for future experiments.

It is important to note that carbon-based materials sputtered directly on the radiator without a~conductive interlayer are sufficient for basic performance studies. Nonetheless, in high-rate environments, a~metallic interlayer is required to mitigate charging-up effects and voltage drops, particularly for larger-area devices.
The influence of the metallic interlayer on the rate capability has not yet been quantified and comparative measurements of carbon-based photocathodes with and without a~Ti layer under high-rate conditions are planned for future studies.

To scale up the detector technology to larger active areas, correspondingly larger photocathodes are required. Concerning the maximum size of the photocathodes, the main limiting factor is the availability of sufficiently flat radiators on the market. The largest commercially available MgF$_2$ substrates are about 10$\times$10 cm$^2$. CsI and DLC depositions on such radiators have been performed, with no uniformity issues observed, as the detectors' response remained uniform across the active area \cite{NDIP,AntonijaMPGD}. The largest PICOSEC Micromegas detector produced to date had an active area of 20$\times$20 cm$^2$, using four individual radiators \cite{MengMPGD}.

Current developments and future perspectives focus on consolidating the detector configuration to further improve stability, including continued investigation of alternative photocathode materials. At the same time, efforts aim to preserve excellent time resolution while meeting additional requirements, such as reducing or replacing CF$_4$~\cite{BrunoldiDRD1}, a~high global warming potential gas, improving spatial resolution~\cite{KallitsopoulouPhDThesis,Kallitsopoulou7pad,BrunbauerSpatial} and ensuring high-rate capability~\cite{GuerraSummer}.

A~recirculation gas system is being under evaluation to reduce gas consumption and enable prolonged operation with minimal or no fresh gas injection. In parallel, studies of alternative CF$_4$-free gas mixtures are ongoing, including Ne:iC$_4$H$_{10}$ and He:iC$_4$H$_{10}$ at different ratios~\cite{BrunoldiDRD1}. Preliminary results are promising for both mixtures, showing performance comparable to the standard gas mixture.

To explore potential improvements in spatial resolution, two sets of resistive prototypes with a~$\varnothing$15 mm active area and increased readout granularity were produced and characterised  using a CsI photocathode with a Ti interlayer ~\cite{BrunbauerSpatial}. Prototypes with a~3.5 mm pad pitch achieved a~spatial resolution of approximately 0.5 mm, with minimal degradation in timing performance at the pad centres.

Another innovative concept investigated to improve spatial resolution was the capacitive-sharing resistive layer architecture \cite{KallitsopoulouPhDThesis,Kallitsopoulou7pad}. This configuration consisted of seven hexagonal pads, each with an outer diameter of approximately 10 mm. The capacitive-sharing prototype, equipped with a CsI photocathode with a Ti interlayer, achieved a time resolution of $\sigma \approx 33$~ps, with spatial resolutions around 1 mm.

A double-layer DLC 100-channel Micromegas with vertical charge evacuation and a~Ti photocathode was studied using a~four-pad scan ~\cite{GuerraSummer}. While the detector achieved a~time resolution of $\mathrm{RMS} \approx 38-41$~ps per pad and $\mathrm{RMS} = 45 \pm 3$~ps for the combined SAT, the spatial resolution, evaluated via a~centre-of-gravity method, was found to be below 3~mm. The observed double-peak structure in the signal arrival time, correlated with the hit position and delayed signals in neighbouring pads, indicated signal propagation effects that could potentially be exploited. Combining these time delays with machine learning approaches may further improve the spatial resolution.

To enable precise timing studies during the Long Shutdown at CERN, a~femtosecond-pulsed laser is being commissioned. Additionally, a new design of a non-bulk PICOSEC Micromegas detector built on the MgF$_2$ radiator to improve pre-amplification gap uniformity is being developed. Further activities include the characterisation of a~UV-sensitive photon-detection prototype. Together, these developments define the next steps toward a~robust and scalable precise-timing detector technology.

Finally, investigations of potential applications of PICOSEC Micromegas detectors in future experiments are ongoing. In particular, applications in future muon collider experiments are under consideration, where the high level of beam-induced background from muon decays requires precise timing for effective event discrimination, with target time resolutions of about 100 ps \cite{MuonCollider}. In addition, the PICOSEC Micromegas detectors' use in the ENUBET/nuSCOPE project \cite{nuSCOPE} for monitored neutrino beams is being explored, including timing at the tagger level for single MIP detection below 100 ps. Test beam campaigns have been performed at CERN using 96-channel detectors equipped with SAMPIC readout \cite{KallitsopoulouPhDThesis,Kallitsopoulou96pad}.
Overall, the studies presented in this paper demonstrate the strong potential of PICOSEC Micromegas detectors as a robust and scalable precise-timing technology for future experiments.

%\acknowledgements
\section*{Acknowledgements}
\label{sec:8}

We acknowledge the support of the CERN EP R\&D Strategic Programme on Technologies for Future Experiments; the DRD1 Collaboration; the PHENIICS Doctoral School Program of Université Paris-Saclay, France; the Cross-Disciplinary Program on Instrumentation and Detection of CEA, the French Alternative Energies and Atomic Energy Commission; the French National Research Agency (ANR), project id ANR-21-CE31-0027; the Program of National Natural Science Foundation of China, grants number 11935014 and 12125505; the COFUND-FP-CERN-2014 program, grant number 665779; the Fundação para a~Ciência e a~Tecnologia (FCT), Portugal; the Enhanced Eurotalents program, PCOFUND-GA-2013-600382; the US CMS program under DOE contract No. DE-AC02-07CH11359. 
%this material is based upon work supported by the U.S. Department of Energy, Office of Science, Office of Nuclear Physics under contracts DE-AC05-06OR23177. 

%\end{linenumbers}

\bibliographystyle{elsarticle-num-names} 
%\bibliography{cas-refs}


\end{document}